\documentclass[preprintnumbers,nofootinbib,secnumarabic]{revtex4}

\usepackage{graphicx}
\usepackage{amsmath,amssymb,mathrsfs}
\usepackage{fancyhdr}
\usepackage{psfrag}
\usepackage{array,bbm}
\usepackage{url}
\usepackage{rotating}
\usepackage{multirow}
\usepackage{dsfont}
\usepackage{cancel}

\numberwithin{equation}{section}
\newcommand*{\CPi}{\ensuremath{\mathrm{CP}_g^{(i)}}}
\newcommand*{\CPii}{\ensuremath{\text{CP}_g^{(ii)}}}

\newcommand*{\trans}{\mathrm{T}}                     
\newcommand*{\unitmatrix}{\mathbbm{1}}
\newcommand*{\id}{\mathbbm{1}_3}
\newcommand*{\tvec}[1]{\ensuremath{\boldsymbol{\mathrm{#1}}}}           

\makeatletter       

\renewcommand{\p@subsection}{}

\makeatother

\DeclareMathOperator{\re}{Re}
\DeclareMathOperator{\im}{Im}

\DeclareMathOperator{\diag}{diag}

\newenvironment{Eqnarray}%
     {\arraycolsep 0.14em\begin{eqnarray}}{\end{eqnarray}}
\newcommand{\beqa}{\begin{Eqnarray}}
\newcommand{\eeqa}{\end{Eqnarray}}
\newcommand{\beq}{\begin{equation}}
\newcommand{\eeq}{\end{equation}}

\def\Eq#1{Eq.~(\ref{#1})}
\def\eq#1{eq.~(\ref{#1})}
\def\eqs#1#2{eqs.~(\ref{#1}) and (\ref{#2})}
\def\Eqs#1#2{Eqs.~(\ref{#1}) and (\ref{#2})}

\def\phm{\phantom{-}}
\def\iso{\mathchoice{\cong}{\cong}{\isoS}{\cong}}
\def\isoS{\vbox{\baselineskip 0pt  \lineskip 0.5pt
    \ialign{$ \mathsurround=0pt  \scriptstyle \hfil ## \hfil $\crcr
        \sim \crcr = \crcr}}}
\def\half{\tfrac{1}{2}}
\def\axis{\boldsymbol{\hat n}}

\newcommand*{\fvec}[1]{\ensuremath{\boldsymbol{\mathrm{\tilde{#1}}}}  }    
\newcommand*{\fmat}[1]{\tilde{#1}}                   

\begin{document}
\preprint{SCIPP 10/15}
\preprint{HD-THEP-10-1}
\title{
Geometric picture of generalized-CP and Higgs-family transformations in the two-Higgs-doublet model}

\author{P.M.~Ferreira}
    \email[E-mail: ]{ferreira@cii.fc.ul.pt}
\affiliation{Instituto Superior de Engenharia de Lisboa,
    Rua Conselheiro Em\'{\i}dio Navarro,
    1900 Lisboa, Portugal}
\affiliation{Centro de F\'{\i}sica Te\'{o}rica e Computacional,
    Faculdade de Ci\^{e}ncias,
    Universidade de Lisboa,
    Av.\ Prof.\ Gama Pinto 2,
    1649-003 Lisboa, Portugal}
\author{Howard E.~Haber}
    \email[E-mail: ]{haber@scipp.ucsc.edu}
\affiliation{Santa Cruz Institute for Particle Physics,
    University of California,
    Santa Cruz, California 95064, USA}
\author{M.~Maniatis}
    \email[E-mail: ]{m.maniatis@thphys.uni-heidelberg.de}
\affiliation{Institut f\"ur Theoretische Physik, Philosophenweg 16, 69120
Heidelberg, Germany}
\author{O.~Nachtmann}
    \email[E-mail: ]{o.nachtmann@thphys.uni-heidelberg.de}
\affiliation{Institut f\"ur Theoretische Physik, Philosophenweg 16, 69120
Heidelberg, Germany}
\author{Jo\~{a}o P.~Silva}
    \email[E-mail: ]{jpsilva@cftp.ist.utl.pt}
\affiliation{Instituto Superior de Engenharia de Lisboa,
    Rua Conselheiro Em\'{\i}dio Navarro,
    1900 Lisboa, Portugal}
\affiliation{Centro de F\'{\i}sica Te\'{o}rica de Part\'{\i}culas,
    Instituto Superior T\'{e}cnico,
    P-1049-001 Lisboa, Portugal}

\date{\today}

\begin{abstract}
In the two-Higgs-doublet model (THDM),
generalized-CP transformations ($\varphi_i\to X_{ij}\varphi_j^*$ where
$X$ is unitary)
and unitary Higgs-family transformations ($\varphi_i\to U_{ij}\varphi_j$)
have recently been examined in a series of papers.
In terms of gauge-invariant bilinear functions of the Higgs fields $\varphi_i$,
the Higgs-family transformations and the generalized-CP transformations
possess a simple geometric description. Namely, these transformations correspond in
the space of scalar-field bilinears to proper and improper rotations,
respectively. In this formalism, recent results relating
generalized CP transformations with Higgs-family transformations
have a clear geometric interpretation.
\end{abstract}

\maketitle

%
\section{Introduction}
\label{sec-intro}

The Standard Model~(SM) of particle physics provides an extremely
successful framework for describing the properties of the fundamental
particles and their interactions.  No statistically significant
deviation from SM predictions has yet been observed in collider
experiments~\cite{pdg}.
Nevertheless, the experimental exploration of the scalar sector of particle physics
is still in its infancy.
The SM contains one complex doublet, hypercharge-one multiplet of scalar Higgs fields.
But it is by no means excluded that the Higgs sector is
larger than that of the SM~\cite{Gunion:1989we}.  In particular, there are a
number of theoretical arguments suggesting a richer
Higgs sector than that of the SM. For instance, the two-Higgs-doublet model~(THDM)
is attractive since it provides a viable framework for spontaneous
CP violation~\cite{Lee:1973iz, Lee:1974jb}.
The Peccei--Quinn symmetry, originally introduced
in order to solve the so-called
{\em strong CP problem}~\cite{Peccei:1977hh, Peccei:1977ur},
requires an enlarged Higgs sector and can be accommodated in the THDM.
Typically, supersymmetric models require at least two Higgs-doublet
fields~\cite{Fayet:1974fj, Fayet:1974pd, Inoue:1982pi, Flores:1982pr, Gunion:1984yn}
in order to cancel potential gauge anomalies due to higgsino superpartners.
Thus, there is ample motivation to study the simplest two-Higgs-doublet extension of the
SM scalar sector.

Indeed, the theory and phenomenology of the THDM has been extensively analyzed;
see~\cite{Deshpande:1977rw,Georgi:1978xz,Haber:1978jt,Donoghue:1978cj,Golowich:1978nh,Hall:1981bc,Gunion:1989we,Haber:1993an,
  Cvetic:1993cy, Botella:1994cs, Lavoura:1994fv, Lavoura:1994yu,  Bernreuther:1998rx,Ginzburg:2004vp, Davidson:2005cw,
Gunion:2005ja, Haber:2006ue, Barbieri:2005kf, Branco:2005em, Barroso:2005sm, Barbieri:2006dq,  Fromme:2006cm,
Barroso:2007rr, Gerard:2007kn, Mahmoudi:2009zx, Ferreira:2010bm,
Velhinho:1994vh, Nagel:2004sw, Ivanov:2005hg, Maniatis:2006jd,Maniatis:2006fs, Nishi:2006tg, Ivanov:2006yq, Ivanov:2007de,
Maniatis:2007vn, Maniatis:2007de, Maniatis:2009vp, Ma:2009ax, Maniatis:2009by}
and references therein.
Among these studies of the THDM, one can find two
lines of approaches.
The traditional approach works directly with the
Higgs-doublet fields~\cite{Deshpande:1977rw,Georgi:1978xz,Haber:1978jt,Donoghue:1978cj,Golowich:1978nh,Hall:1981bc,Gunion:1989we,Haber:1993an,
Cvetic:1993cy, Botella:1994cs, Lavoura:1994fv,
  Lavoura:1994yu,  Bernreuther:1998rx,Ginzburg:2004vp,
Davidson:2005cw, Gunion:2005ja, Haber:2006ue, Barbieri:2005kf, Branco:2005em, Barroso:2005sm, Barbieri:2006dq,  Fromme:2006cm,
Barroso:2007rr, Gerard:2007kn, Mahmoudi:2009zx, Ferreira:2010bm}.
In contrast, there is a second approach that emphasizes the
role of gauge-invariant scalar field bilinears~\cite{
Velhinho:1994vh, Nagel:2004sw, Ivanov:2005hg, Maniatis:2006jd,Maniatis:2006fs, Nishi:2006tg, Ivanov:2006yq, Ivanov:2007de,
Maniatis:2007vn, Maniatis:2007de, Maniatis:2009vp, Ma:2009ax, Maniatis:2009by}.
A systematic use of the scalar field bilinears for the study of the
stability and the structure of electroweak symmetry breaking~(EWSB) in the THDM
was first carried out in~\cite{Nagel:2004sw,Maniatis:2006fs}.
Then independently in~\cite{Nishi:2006tg} the scalar field
bilinears were introduced and employed for the study
of the CP properties of the THDM.
This latter approach was revisited in~\cite{Ivanov:2006yq}.
In a recent paper~\cite{Ferreira:2009wh}, three of the present
authors mainly employed the traditional approach in the study of Higgs-family and
generalized-CP (GCP) symmetries of the THDM.
It is the purpose of this work to present a compelling geometrical interpretation of
Higgs-family and GCP symmetries using the formalism of scalar field bilinears.
This formalism provides a powerful geometric framework that yields new insights and clarifies the relations among
the different possible symmetry transformations of the THDM.

In Section 2, we review the formalism of scalar field bilinears and their applications in the
analysis of the THDM.  In Section 3, we introduce the generalized-CP (GCP) transformations.
These transformations are initially defined as transformations of scalar fields.  We then
revisit the GCP transformations in the formalism of scalar field bilinears and provide
a geometric interpretation.  Two useful theorems involving GCP transformations are proven
in Section~4.  Higgs-family and GCP symmetries are examined in detail
in Section 5.  The classification
of all possible THDM symmetries is established in the formalism of scalar field bilinears.
The constraints on the scalar potential parameters due to the various possible symmetry classes
is provided in a form that is covariant with respect to arbitrary transformations of
the basis for the scalar fields.  A distinction between parameter constraints derived
in an arbitrary basis and in a specific basis is examined and clarified.  Conclusions are
give in Section 6.  Details on the structure of $3\times 3$ rotation matrices that
are useful for the computations performed in this paper are provided in Appendix A.

%
\section{Scalar field bilinears in the THDM}
\label{secII}

The scalar sector of the THDM contains two complex doublet,
hypercharge-one Higgs fields,
with respect to the SU(2)$\times$U(1) electroweak gauge group, denoted by
\begin{equation}
\varphi_i(x) =
\begin{pmatrix}
\varphi_i^+(x)\\
\varphi_i^0(x)
\end{pmatrix}\;,
\qquad
\text{with }i=1,2\;.
\end{equation}
The tree-level THDM Lagrangian contains the kinetic term and
the potential $V(\varphi_1,\varphi_2)$ for the Higgs fields, which is
gauge invariant and renormalizable.  These requirements imply that the scalar
potential  $V(\varphi_1,\varphi_2)$ consists only of
quadratic and quartic terms
in the fields. The conventional parametrization in the field
approach reads~\cite{Gunion:1989we, Haber:1993an}
\begin{equation}
\label{eq5}
\begin{split}
V =~&
m_{11}^2 (\varphi_1^\dagger \varphi_1) +
m_{22}^2 (\varphi_2^\dagger \varphi_2) -
m_{12}^2 (\varphi_1^\dagger \varphi_2) -
(m_{12}^2)^* (\varphi_2^\dagger \varphi_1)\\
& +\tfrac{1}{2} \lambda_1 (\varphi_1^\dagger \varphi_1)^2
+ \tfrac{1}{2} \lambda_2 (\varphi_2^\dagger \varphi_2)^2
+ \lambda_3 (\varphi_1^\dagger \varphi_1)(\varphi_2^\dagger \varphi_2) \\
&+ \lambda_4 (\varphi_1^\dagger \varphi_2)(\varphi_2^\dagger \varphi_1)
+ \tfrac{1}{2} [\lambda_5 (\varphi_1^\dagger \varphi_2)^2 + \lambda_5^*
(\varphi_2^\dagger \varphi_1)^2] \\
&+ [\lambda_6 (\varphi_1^\dagger \varphi_2) + \lambda_6^*
(\varphi_2^\dagger \varphi_1)] (\varphi_1^\dagger \varphi_1) + [\lambda_7 (\varphi_1^\dagger
\varphi_2) + \lambda_7^* (\varphi_2^\dagger \varphi_1)] (\varphi_2^\dagger \varphi_2)\;,
\end{split}
\end{equation}
with $m_{11}^2$, $m_{22}^2$, $\lambda_{1,2,3,4}$ real and
$m_{12}^2$, $\lambda_{5,6,7}$
potentially complex.
The parameters of the scalar potential must be chosen such that the
potential is bounded from below, is stable, and leads to the correct
form for EWSB [which preserves U(1)$_{\rm EM}$].
In addition, one may choose to impose various additional symmetry
requirements on the scalar potential.

A very convenient way to study the stability, the structure of the EWSB, and
any additional symmetry requirements is to use scalar field
bilinears~\cite{Nagel:2004sw,Ivanov:2005hg,Maniatis:2006fs,Nishi:2006tg,Ivanov:2006yq,Ivanov:2007de}.
We follow here the notation of~\cite{Nagel:2004sw,Maniatis:2006fs}
and define the four
independent gauge invariant bilinears as
\begin{equation}
\label{eq9}
\begin{aligned}
K_0 &= \varphi_1^\dagger \varphi_1 + \varphi_2^\dagger \varphi_2, &
K_1 &= \varphi_1^\dagger \varphi_2 + \varphi_2^\dagger \varphi_1,\\
K_2 &= i \varphi_2^\dagger \varphi_1 -  i \varphi_1^\dagger \varphi_2, &
K_3 &= \varphi_1^\dagger \varphi_1 - \varphi_2^\dagger \varphi_2.
\end{aligned}
\end{equation}
We summarize some results from~\cite{Nagel:2004sw,Maniatis:2006fs}.
\begin{itemize}
\item We have
\begin{equation}
K_0 \ge 0\;, \qquad K_0^2 - K_1^2 -K_2^2-K_3^2 \ge 0\;.
\end{equation}
That is, the four vectors
$\fvec{K} =(K_0, \tvec{K})^\trans = (K_0, K_1, K_2, K_3)^\trans$ span the
forward light cone in $K$ space. These $\fvec{K}$
parameterize the gauge orbits of the Higgs-doublet fields.

\item A change of basis of the Higgs fields, called a
Higgs-family transformation
\begin{equation}
\label{eq11}
\begin{pmatrix} \varphi_1 \\
                \varphi_2 \end{pmatrix}
\to
\begin{pmatrix} \varphi'_1 \\
                \varphi'_2 \end{pmatrix}
= U
  \begin{pmatrix} \varphi_1 \\
                  \varphi_2 \end{pmatrix}\;,
\end{equation}
with $U= (U_{ij}) \in {\rm U}(2)$, corresponds
to an ${\rm SO}(3)$ rotation in $K$ space
\begin{equation}
\label{eq12}
\begin{split}
&K_0 \to K'_0 = K_0,\\
&\tvec{K} \to \tvec{K}' = R(U)  \tvec{K}\;.
\end{split}
\end{equation}
Here, the $3\times 3$ real orthogonal matrix $R(U)$ is obtained from
\begin{equation}
\label{eq13}
U^\dagger \sigma^a U = R_{ab}(U)\,\sigma^b.
\end{equation}
Since $U$ is continuously connected to the identity,
it follows that ${\rm det}~R(U)=1$.
It is straightforward to prove that
\beq \label{groupmult}
R_{ab}(U)R_{bc}(V)=R_{ac}(UV)\,,
\eeq
starting from \eq{eq13} and using the fact that the $\sigma^a$
span the space of traceless $2\times 2$ Hermitian matrices.
Thus, the mapping $\{U\,,\,-U\}\longmapsto R(U)$ provides the
well-known double
cover of SO(3) by SU(2).  An explicit formula for $R_{ab}(U)$ is easily
obtained:
\beq \label{Rabu}
R_{ab}(U)=\tfrac{1}{2}{\rm Tr}(U^\dagger\sigma^a U \sigma^b)\,.
\eeq
The most general SO(3) matrix can be uniquely specified by an axis of
rotation $\axis$ and an angle of rotation $\theta$ that lies in the interval
$0\leq\theta\leq\pi$.  We denote the corresponding $3\times 3$ matrix by
$R(\axis,\theta)$.  The properties of $R(\axis,\theta)$ are reviewed in
Appendix A.

\item Every ${\rm SO}(3)$ rotation in $K$ space, given by~\eqref{eq12}
with $R \in {\rm SO}(3)$, corresponds to
a Higgs-family transformation~\eqref{eq11}
that is unique up to gauge transformations.

\item The most general gauge invariant and
renormalizable potential~$V$ of the
THDM can be written as
\begin{equation}
\label{eq-Vfour}
V= \fvec{K}^\trans \fvec{\xi} + \fvec{K}^\trans \fmat{E} \fvec{K}\,.
\end{equation}
Here, $\fvec{\xi}$ and $\fmat{E}$ contain the parameters
which are all real,
\begin{equation}
\label{eq-fourpar}
\fvec{\xi} =
\begin{pmatrix}
\xi_0\\
\tvec{\xi}
\end{pmatrix},
\qquad
\fmat{E} =
\begin{pmatrix}
\eta_{00} & \,\,\,\tvec{\eta}^\trans\\
\tvec{\eta} & \,\,\, E
\end{pmatrix}\,,
\end{equation}
with $E=E^\trans$ a $3\times 3$~matrix.
Expressing $\tilde{\tvec{\xi}}$ and $\tilde{\tvec{E}}$ in terms
of the parameters of~\eqref{eq5}, we find
\begin{equation}
\label{eq17}
\begin{aligned}
&\xi_0=\tfrac{1}{2}
(m_{11}^2+m_{22}^2)\;,
\qquad
\tvec{\xi}=\tfrac{1}{2}
\begin{pmatrix}
- 2 \re(m_{12}^2)\\
\phantom{+} 2 \im(m_{12}^2)\\
 m_{11}^2-m_{22}^2
\end{pmatrix}\;,
\qquad
\tvec{\eta}=\tfrac{1}{4}
\begin{pmatrix}
\phantom{+}\re(\lambda_6+\lambda_7)\\
-\im(\lambda_6+\lambda_7)\\
\tfrac{1}{2}(\lambda_1 - \lambda_2)
\end{pmatrix}\;,\\
&\eta_{00} =
\tfrac{1}{8}(\lambda_1 + \lambda_2) + \tfrac{1}{4}\lambda_3\;,
\qquad
E = \tfrac{1}{4}
\begin{pmatrix}
\lambda_4 + \re(\lambda_5) &
-\im(\lambda_5) &
\re(\lambda_6-\lambda_7) \\
-\im(\lambda_5) &
\lambda_4 - \re(\lambda_5) &
-\im(\lambda_6-\lambda_7) \\
\re(\lambda_6-\lambda_7) &
-\im(\lambda_6 -\lambda_7) &
\tfrac{1}{2}(\lambda_1 + \lambda_2) - \lambda_3
\end{pmatrix}.
\end{aligned}
\end{equation}

\item A transformation~\eqref{eq11} corresponds
to a {\em Higgs family symmetry} if and only if
\begin{equation}
\label{eq18}
R(U) \tvec{\xi} = \tvec{\xi}\;, \qquad
R(U) \tvec{\eta} = \tvec{\eta}\;, \qquad
R(U) E R^\trans (U) = E\;.
\end{equation}
\end{itemize}

In~\cite{Ivanov:2006yq} basis changes of the Higgs fields as
in~\eq{eq11} were considered, but with the unitary transformation $U$
replaced by more general ${\rm SL}(2,\mathbbm{C})$
transformations. In $K$ space this corresponds to general
Lorentz transformations, which includes both rotations and boosts.
However, the latter change the form of the kinetic terms of the THDM Lagrangian.
Without loss of generality,
one may assume that the kinetic terms of the tree-level
THDM Lagrangian are of canonical form.  Under this assumption,
only unitary Higgs family transformations are permitted.

%
\section{Generalized-CP transformations}
\label{secIV}

In this section we study the generalized-CP~(GCP) transformations,
\begin{equation}
\label{eq27}
\varphi_i(x) \rightarrow X_{ij} \; \varphi_j^*(x'),\quad \text{with }
i,j \in \{1,2\} \;, \quad X = \left(X_{ij} \right) \in {\rm U}(2), \quad
x=\begin{pmatrix} x^0 \\ \tvec{x} \end{pmatrix},\;
x'=\begin{pmatrix} x^0 \\ -\tvec{x} \end{pmatrix}.
\end{equation}
Such GCP transformations of scalar fields have been previously considered
in~\cite{Ecker:1981wv,Ecker:1983hz,Ecker:1987qp,Neufeld:1987wa,Branco:1999fs}
\footnote{For the discussion of CP transformations in the SM see for
example~\cite{Nachtmann:1990ta, Branco:1999fs, bigi}.},
and GCP transformations of fermions
fields have also been examined in~\cite{Bernabeu:1986fc}. A systematic study of
GCP transformations of the scalar fields of the THDM was carried out
in~\cite{Maniatis:2007vn} and~\cite{Ferreira:2009wh}.
The matrix $X$ that appears in \eq{eq27} is basis-dependent.  Under a change of basis specified
by \eq{eq11}, the GCP transformation of \eq{eq27} is modified to:
\beq
\varphi^\prime_i(x) \rightarrow X^\prime_{ij} \; \varphi_j^{\prime\,*}(x')\,,
\eeq
where $X^\prime$ is a unitary matrix given by:
\beq \label{ubasis}
X^\prime\equiv UXU^\trans\,.
\eeq

As shown in \cite{Ferreira:2009wh}, three classes of GCP transformations exist depending on the value of (GCP)$^2$.
Consider first the case of (GCP)$^2=1$, which is denoted by CP1
(sometimes called the ``standard'' CP symmetry transformation).
Then,
\beq
\varphi_i\to X_{ij}\varphi_j^{*}\to X_{ij}X_{jk}^*\varphi_k
=\varphi_i\,,
\eeq
which implies that $XX^*=\mathds{1}_2$ (where $\mathds{1}_2$ is the $2\times 2$ identity matrix).
Since $X$ is unitary, the latter implies
that $X$ is also symmetric.  Thus, \eq{eq27} corresponds to a CP1
transformation if and only if $X$ is a symmetric unitary matrix.
One can now employ the well known result that
any symmetric unitary matrix $X$ can be written as the product
of a unitary matrix and its transpose
(see e.g. Appendix D.3 of \cite{Dreiner:2008tw} for a proof of this result).
That is, one can always find a unitary matrix $U$ such that
$X=U^\dagger U^*$.  Performing the basis transformation given by
\eq{ubasis} then yields
that
\beq \label{res1}
X^\prime\equiv UXU^\trans=UU^\dagger (UU^\dagger)^*=\mathds{1}_2\,.
\eeq
That is, in the case of CP1, there is always a basis choice for which
$X^\prime=\mathds{1}_2$.

Next, consider the case of (GCP)$^2=-1$, which is denoted by CP2.
Then,
\beq
\varphi_i\to X_{ij}\varphi_j^{*}\to X_{ij}X_{jk}^*\varphi_k
=-\varphi_i\,,
\eeq
which implies that $XX^*=-\mathds{1}_2$.  Since $X$ is unitary, the latter implies
that $X$ is also antisymmetric.  Thus, \eq{eq27} corresponds to a CP2
transformation if and only if $X$ is an antisymmetric unitary matrix.
The most general antisymmetric unitary $2\times 2$ matrix
$X$ is\footnote{Conversely, it is straightforward to
show that if $X^\dagger\sigma^a X=-\sigma^{a*}$, where $X=\exp(i\theta
{\boldsymbol{\hat n\cdot\vec\sigma}}/2)$, then $X$ is proportional to
  $\sigma^2$, i.e. $X$ is an antisymmetric unitary matrix.}
\beq \label{epsdef}
X=e^{i\psi}\epsilon\,,\qquad \text{where}~~\epsilon\equiv i\sigma^2=-\epsilon^\trans=-\epsilon^{-1}=
\left(\begin{matrix} \phm 0 & \,\, 1 \\  -1 & \,\, 0
\end{matrix}\right)\,,
\eeq
where $\psi$ is an arbitrary phase.  It then follows that a unitary
matrix exists, $U\equiv e^{-i\psi/2}\epsilon$, such that
\beq \label{res2}
X^\prime=UX U^\trans=\epsilon\,.
\eeq
That is,
in the case of CP2, there is always a basis choice for which
$X^\prime=\epsilon$.

Finally, we consider the case of (GCP)$^2=XX^*\neq \pm \mathds{1}_2$,
which is denoted by CP3.  In this case, it is always possible to
perform a basis transformation such that in the new basis, $X$ is
transformed into:\footnote{\Eq{UXUT} is an example of a canonical form for
unitary congruence.  For a comprehensive mathematical treatment,
see \cite{horn} (note in particular Corollary 8.7).
In the physics literature, \eq{UXUT} first appeared
in \cite{Ecker:1987qp} and was further generalized in \cite{Grimus:1988qr}.}
\beq \label{UXUT}
X^\prime =UXU^\trans=\begin{pmatrix} \phm\cos \theta & \quad \sin \theta  \\
-\sin \theta & \quad \cos \theta \end{pmatrix}\,,
\eeq
where $0<\theta<\pi/2$.

To prove this result, we first note that since $X$ is a unitary matrix,
${\rm det}~X\equiv e^{2i\chi}$ is a pure phase.  Following
\eq{ubasis}, we shall perform a
basis transformation such that
\beq \label{detX}
{\rm det}~X^\prime={\rm det}(UXU^\trans)=1\,.
\eeq
This can always be done provided that
\beq \label{detU}
U=e^{-i\chi/2}\widehat U\,,
\eeq
where $\widehat U$ is an SU(2) matrix.  It is convenient to define:
\beq
\widehat X\equiv e^{-i\chi}X\,,
\eeq
in which case ${\rm det}~\widehat X=1$ and
\beq \label{xp}
X^\prime=\widehat U\widehat X\widehat U^\trans\,.
\eeq
A general SU(2) matrix $\widehat U$ satisfies:
\beq \label{real}
\widehat U=\epsilon \,\widehat U^*\epsilon^{-1}\,,
\eeq
where $\epsilon$ is
defined in \eq{epsdef}.  \Eq{real} expresses the well known
equivalence of the irreducible two-dimensional representation of SU(2) and its
complex conjugate.  Inserting the transpose of \eq{real}
into \eq{xp} yields:
\beq \label{Xprime}
X^\prime\epsilon=\widehat U\widehat X\epsilon\, \widehat U^\dagger\,.
\eeq

It is convenient to define:
\beq \label{V}
\widehat U\equiv\frac{1}{\sqrt{2}}
\begin{pmatrix} 1 & \quad \phm 1 \\ i &
  \quad -i \end{pmatrix} V\,,
\eeq
where $V$ is a unitary matrix (such that ${\rm det}~V=i$).  Since $\widehat X\epsilon$ is an SU(2)
matrix, it follows that the two eigenvalues of $\widehat X\epsilon$
are complex conjugates of each other, denoted below by $e^{\pm i\phi}$,
where the real angle $\phi$ is defined modulo $\pi$.  Then, we can
choose $V$ to be the unitary matrix that diagonalizes $\widehat X\epsilon$,
\beq \label{diag}
V\widehat X\epsilon\, V^\dagger=
\begin{pmatrix} e^{i\phi} & \quad 0 \\ 0 & \quad e^{-i\phi}
\end{pmatrix}\,.
\eeq
Inserting \eqs{V}{diag} into \eq{Xprime} yields
\beq
X^\prime=\begin{pmatrix} \sin\phi & \quad -\cos\phi  \\
\cos\phi & \quad \phm\sin\phi \end{pmatrix}\,.
\eeq
Finally, we define $\phi=\theta+\pi/2$ to obtain the desired form
given by \eq{UXUT}.

The angle $\phi$ (and hence the angle $\theta$) is defined modulo $\pi$.
Thus, it is convenient to establish a convention in which $|\theta|\leq\pi/2$.
However, we are free to redefine $U\to \sigma^1 U$,
which has the effect of changing
the overall sign of~$\theta$.\footnote{For any
$2\times 2$ matrix $A$, the matrix $\sigma^1 A\sigma^1$ is related
to $A$ by an
interchange of the two diagonal elements and an interchange of the two
off-diagonal elements.}
Consequently, it is always possible to find a basis transformation
such that $X^\prime$ takes the form given by \eq{UXUT}, where
$0\leq\theta\leq\pi/2$.  Moreover,
\beq
XX^*=U^\dagger X^\prime U^*U^\trans X^{\prime\,*}U=
U^\dagger \begin{pmatrix} \phm\cos 2\theta & \quad \sin 2\theta  \\
-\sin 2\theta & \quad \cos 2\theta \end{pmatrix}U\,,
\eeq
so that
$\theta=0$ corresponds to the case of CP1 [\eq{res1}], $\theta=\pi/2$
corresponds to the case of CP2 [\eq{res2}], and $0<\theta<\pi/2$
corresponds to the case of CP3 [\eq{UXUT}].

Summarizing the above results,
it follows that in
a suitable basis for the scalar fields
the matrix $X$ in~\eqref{eq27} can
always be brought
to the form
\begin{equation}
\label{eq32}
\begin{pmatrix}
\phm\cos \theta &\quad \sin \theta \\
-\sin \theta & \quad \cos \theta
\end{pmatrix},
\qquad \text{with } 0 \le \theta \le \pi/2 .
\end{equation}
The classification of GCP symmetries established above is~\cite{Ferreira:2009wh}:
\begin{itemize}
\item CP1 if $\theta =0$,
\item CP2 if $\theta = \pi/2$ and
\item CP3 if $0< \theta < \pi/2$.
\end{itemize}

We now demonstrate how the classification of GCP symmetries can
be understood in the formalism of field bilinears employed in
\cite{Maniatis:2007vn}.  To make the present paper self-contained, we
shall repeat some of the derivations of \cite{Maniatis:2007vn}
in the analysis that follows.

In the notation of \eq{eq13},  we define
the SO(3) matrix $R_{ab}(X)$ via:
\begin{equation}
\label{RX}
X^\dagger \sigma^a X = R_{ab}(X)\,\sigma^b\,,
\end{equation}
where $X$ is the unitary matrix that specifies the GCP transformation
[cf.~\eq{eq27}]. It is convenient to introduce the improper rotation
matrix,
\beq
\overline R_2\equiv {\rm diag}(1,-1,1)\,.
\eeq
The matrix $\overline{R}_2$ describes the reflection through the
1--3~plane in $K$ space.
Similarly, we introduce $\overline{R}_1$ and $\overline{R}_3$ as
the reflections through the 2--3 and 1--2 planes,
respectively,\footnote{Here and in the following $R$, $R_\alpha,\ldots$
and $\overline{R}$, $\overline{R}_\alpha,\ldots$ denote
proper and improper rotation matrices with determinant $+1$ and $-1$,
respectively.}
\begin{equation}
\label{eq31}
\begin{split}
\overline{R}_1 &= \diag(-1,1,1) ,\\
\overline{R}_3 &= \diag(1,1,-1).
\end{split}
\end{equation}
In particular, note that
\beq \label{sigR}
\sigma^{a *}=\sigma^{a\,T}=(\overline R_2)_{ab}\sigma^b\,.
\eeq

The scalar field bilinears of \eq{eq9} can be rewritten as:
\beq
K_\mu=\varphi_i^\dagger\sigma^\mu_{ij}\varphi_j\,,
\eeq
where $\sigma^\mu=(1,\boldsymbol{\vec\sigma})$.
Then, the GCP transformation defined in \eq{eq27} corresponds to
\beq
K_\mu=\varphi_i^\dagger\sigma^\mu_{ij}\varphi_j \to
(X_{ik}\varphi_k^\dagger)^*\sigma^\mu_{ij}X_{j\ell}\varphi^*_\ell
= \varphi^\dagger_\ell\varphi_k (X^\dagger \sigma^\mu X)_{k\ell}\,.
\eeq
If $\mu=0$, then $X^\dagger\sigma^0 X=X^\dagger X=\mathds{1}_2$, and so
$K_0\to K_0$ (where we suppress the coordinates $x$ and $x'$).  If $\mu=a=1,2,3$,
then one may use \eqs{RX}{sigR} to obtain:
\beqa
K_a &\to & \varphi^\dagger_\ell\varphi_k (X^\dagger \sigma^a X)_{k\ell}
\nonumber \\
&=&  \varphi^\dagger_\ell\varphi_k R_{ab}(X)\sigma^b_{k\ell}\nonumber  \\
&=& \varphi^\dagger_\ell\varphi_k R_{ab}(X)
(\overline{R}_2)_{bc}\sigma^c_{\ell k} \nonumber \\
&=&\overline{R}_{ac}K_c\,,
\eeqa
where  $\overline{R}$ is the improper rotation matrix:
\begin{equation}
\label{eq30}
\overline{R} \equiv R(X) \overline{R}_2\,.
\end{equation}
That is,
\begin{equation}
\label{eq29}
\begin{split}
K_0(x)  &\rightarrow K_0(x'),\\
\tvec{K}(x) &\rightarrow \overline{R}\; \tvec{K}(x')\,,
\end{split}
\end{equation}
which reproduces the result obtained in section 3 of~\cite{Maniatis:2007vn}.

Under each of the three classes of GCP transformations,
the improper rotation matrix $\overline{R}\equiv R(X)\overline{R}_2$
must satisfy an appropriate constraint equation.  To derive the
relevant constraint, we start with the complex conjugate of \eq{RX}.
Using \eq{sigR}, it then follows that
\beq
X^\trans(\overline{R}_2)_{ab}\sigma^b X^*=R_{ab}(X)(\overline{R}_2)_{bc}\sigma^c\,.
\eeq
Employing \eq{RX} once more yields
\beq \label{bRR}
(\overline{R}_2)_{ab}R_{bc}(X^*)\sigma^c
=R_{ab}(X)(\overline{R}_2)_{bc}\sigma^c\,.
\eeq
Since the $\sigma^a$ are linearly independent and
span the space of traceless $2\times 2$
Hermitian matrices, \eq{bRR} yields:
\beq \label{RXs}
R(X^*)=\overline{R}_2 R(X) \overline{R}_2\,,
\eeq
after using $(\overline{R}_2)^2=\mathds{1}_3$ (where
$\mathds{1}_3$ is the $3\times 3$ identity matrix).
Finally, we multiply \eq{RXs} on the left by $R(X)$ and make use of
\eq{groupmult} to obtain:
\beq \label{rw}
R(XX^*)=\overline{R}^2\,,
\eeq
where the improper rotation matrix $\overline{R}\equiv
R(X)\overline{R}_2$ was introduced in \eq{eq30}.

Consider separately the cases CP2 (where $X$ is antisymmetric and
$XX^*=-\mathds{1}_2$) and CP1 (where $X$ is symmetric and $XX^*=\mathds{1}_2$).
In both cases, \eq{RX} yields
$R(XX^*)=R(\pm\mathds{1}_2)=\mathds{1}_3$, and it follows
that $\overline{R}^2=\mathds{1}_3$.  That is,
$\overline{R}$ is either a reflection matrix corresponding to
a reflection through some plane in $K$ space or an
inversion (or point reflection) through the origin in $K$ space.
In \eq{epsdef}, we noted the most general form
for $X$ in the case of CP2 is given by $X=e^{i\psi} i \sigma^2$,
where $\psi$ is an arbitrary phase.  Inserting this result
into \eq{RX} and making use of \eq{sigR} yields:
\beq
R_{ab}(X) \sigma^b = \sigma^2\sigma^a\sigma^2=-\sigma^{a*}=-(\overline R_2)_{ab}\sigma^b\,,
\eeq
and we conclude that $R(X)=-\overline{R}_2$.
Hence, \eq{eq30} yields $\overline{R}=-\mathds{1}_3$,
which corresponds to an inversion
(i.e., a point reflection through the origin in $K$ space). This case is
a CP$^{(i)}_g$ transformation in the notation of
\cite{Maniatis:2007vn}.
In contrast, for the case of CP1, where $X$ is a symmetric
unitary matrix, we have $\overline{R}^2=\mathds{1}_3$
and $\overline{R}\neq -\mathds{1}_3$.  An
improper rotation matrix of this type corresponds to reflections through planes in
$K$ space.  In particular, an SO(3) matrix $\widetilde R$ exists such
that
\beq \label{RR}
\overline{R}=\widetilde{R} \overline{R}_2 \widetilde{R}^\trans\,.
\eeq
To prove \eq{RR}, simply choose a basis where $X=\mathds{1}_2$
in which case
$R(X)=\mathds{1}_3$ and $\overline{R}=\overline{R}_2$.  Then, rotate in $K$ space
to an arbitrary basis using the rotation matrix $\widetilde {R}$ to
obtain \eq{RR}.
One can easily check that $\overline{R}^2=\mathds{1}_3$ and $\overline{R}\neq
-\mathds{1}_3$ as required.  This case corresponds to
a CP$^{(ii)}_g$ transformation
in the notation of \cite{Maniatis:2007vn}.  In summary, for those
GCP transformations whose square is equal to the unit transformation
when acting on the gauge invariant field bilinears, we must have
$\overline{R}^2=\mathds{1}_3$.  The resulting classification of
\cite{Maniatis:2007vn} is then related to that of \cite{Ferreira:2009wh}
as follows:
\begin{itemize}
\item \CPii : reflections through planes in $K$ space,
$\overline{R} = \tilde{R} \overline{R}_2 \tilde{R}^\trans$ with
$\tilde{R} \in {\rm SO}(3)\quad\Longleftrightarrow\quad
{\rm CP1}$~transformations.
\item \CPi : a point reflection through the origin
in $K$ space, $\overline{R} = -\unitmatrix_3\quad\Longleftrightarrow\quad
{\rm CP2}$~transformations.
\end{itemize}

The case of CP3 transformations was not considered in detail in \cite{Maniatis:2007vn}.
In this case, $XX^*\neq \pm\mathds{1}_2$, which implies that
$R(XX^*)\neq \mathds{1}_3$.  Hence, \eq{rw} yields
$\overline{R}^2\neq \mathds{1}_3$, which means that the improper rotation $\overline{R}$ is
\textit{not} a reflection matrix or an inversion.  To obtain an explicit form for
$\overline{R}\equiv R(X)\overline{R}_2$, it is convenient to choose a suitable basis in which
$X$ is given by \eq{eq32}, which can be rewritten as:
\beq \label{Xspecial}
X=\mathds{1}_2\cos\theta+i\sigma^2\sin\theta\,.
\eeq
We can use \eq{Rabu} to obtain the corresponding rotation matrix $R_{ab}(X)$.
Evaluating the relevant traces, one obtains:
\beq \label{Rabx}
R_{ab}(X)=\delta_{ab}\cos 2\theta+2\delta_{a2}\delta_{b2}\sin^2\theta+\epsilon_{ab2}\sin 2\theta\,.
\eeq
Hence, in a basis in which $X$ is given by \eq{eq32},
\beq \label{eq35}
\overline{R}\equiv R(X)\overline{R}_2=\begin{pmatrix} \cos 2\theta & \quad \phm 0 & \quad -\sin 2\theta \\
0 & \quad -1 & \quad 0 \\ \sin 2\theta & \quad \phm 0 & \quad \phm\cos 2\theta \end{pmatrix}\,.
\eeq
As expected, for a CP1 transformation $\theta=0$ and
$\overline{R}=\overline{R}_2$ [in a suitable basis in which
$X=\mathds{1}_2$], and for a CP2 transformation $\theta=\pi/2$ and
$\overline{R}=-\mathds{1}_3$.  The case of $0<\theta<\pi/2$
corresponds to a CP3 transformation, in which $\overline{R}$ is an
improper rotation matrix that is not a simple reflection or inversion
(i.e.~$\overline{R}^2\neq \mathds{1}_3$).  As in \eq{RR}, the
form for $\overline{R}$ in a general basis is related to
\eq{eq35} by an orthogonal similarity transformation,
\beq
\overline{R}=\widetilde R R(X) \overline{R}_2\widetilde R^\trans\,,
\eeq
for some SO(3) matrix $\widetilde R$,
where $R(X)$ is given by \eq{Rabx}.

Thus, we have reproduced above the result proved in~\cite{Maniatis:2007vn}.
\textit{Every  transformation of the field bilinears given by~\eq{eq29},
where $\overline{R}$ is \textit{any} improper
rotation matrix, corresponds to a
GCP transformation of the
fields as specified in~\eq{eq27}.
}
This field transformation is
uniquely determined by $\overline{R}$
up to gauge transformations.
In analogy with \eq{eq18},
a GCP transformation of the form given by \eq{eq29} corresponds to
a \textit{GCP symmetry} if and only if
\begin{equation}
\label{GCPsymm}
\overline{R} \tvec{\xi} = \tvec{\xi}\;, \qquad
\overline{R} \tvec{\eta} = \tvec{\eta}\;, \qquad
\overline{R} E \overline{R}^\trans = E\;.
\end{equation}

%
\section{Two Theorems involving GCP transformations}
\label{secIVa}

In \cite{Ferreira:2009wh}, it was suggested that all symmetries of the THDM
could be expressed in terms of products of GCP symmetries.  In
this section, we prove two simple theorems that demonstrate that
all Higgs family and GCP
transformations can be expressed in terms of products of CP1 transformations.
\vskip 0.1in

\noindent
\textbf{Theorem 1:} Any Higgs-family transformation
can be considered as a product
of two CP1 transformations.
\vskip 0.1in

\noindent
\textbf{Theorem 2:} Any GCP transformation
is either a CP1 transformation or a
product of three CP1 transformations.
\vskip 0.1in

\noindent
These theorems are new. Some particular cases were
considered in~\cite{Ferreira:2009wh} but even there they referred
only to \text{symmetries} where the form of the
potential could play a role.
The results presented here apply to the more fundamental
\textit{transformations} themselves.
\vskip 0.1in

\textbf{Proof of Theorem~1}.   Let $X$ and $Y$ be two
arbitrary symmetric unitary matrices.  Consider the product of the
corresponding CP1 transformations,
\beqa
\varphi^\prime_j(x)&=&X_{jk}\varphi_k^*(x')\,,\\
\varphi_i^{\prime\prime}(x)&=&Y_{ij}\varphi_j^{\prime\,*}(x')=Y_{ij}X_{jk}^*\varphi_k(x)=
U_{ik}\varphi_k(x)\,,
\eeqa
where $U=YX^*$ is a unitary matrix.  The theorem is proven if we can
show that an arbitrary unitary matrix $U$ is the product of two
symmetric unitary matrices.  But this last statement is easy to prove.
First, we diagonalize $U$ with a unitary matrix $W$,
\beq
U=WDW^\dagger\,,
\eeq
where $D$ is a diagonal matrix of phases corresponding to the
eigenvalues of $U$.  Then define the following two symmetric unitary
matrices:
\beq
S_1\equiv WW^\trans\,,\qquad S_2\equiv W^*DW^\dagger\,.
\eeq
It immediately follows that
\beq \label{ss}
S_1 S_2 = WW^\trans W^* DW^\dagger=WDW^\dagger=U\,,
\eeq
which shows that any unitary matrix can be written as a product of two
symmetric unitary matrices.

\begin{figure}[ht!]
\includegraphics[width=0.5\textwidth]{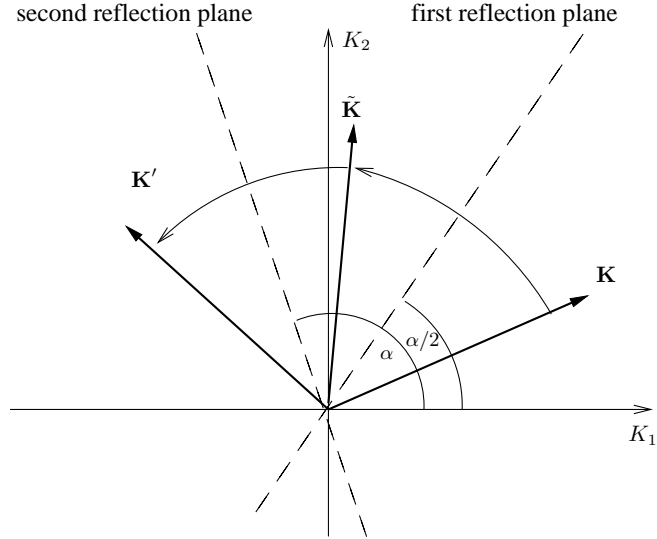}
\caption{\label{fig1}
The reflections $\overline{R}_{\alpha/2}$ and
$\overline{R}_\alpha$ illustrated in the
$K_1$--$K_2$ plane. The reflection
$\overline{R}_{\alpha/2}$ brings the arbitrary
vector~$\tvec{K}$ to $\tilde{\tvec{K}}$.
Then, $\overline{R}_{\alpha}$ brings
$\tilde{\tvec{K}}$ to $\tvec{K}'$.}
\end{figure}

It is also instructive to prove Theorem 1 in $K$ space.
Consider an arbitrary
Higgs-family transformation [cf.~\eqs{eq11}{eq12}].
Every proper rotation matrix~$R$ is a
rotation about an axis and can be represented,
in a suitable basis, as
\begin{equation}
\label{eq37}
R_\alpha=
\begin{pmatrix}
\cos \alpha & \quad -\sin \alpha & \quad 0\\
\sin \alpha & \quad \phm\cos \alpha & \quad 0\\
0 & \quad 0 & \quad 1\\
\end{pmatrix},
\qquad \text{with } 0 \le \alpha \le \pi.
\end{equation}
This rotation can also be generated by two
reflections, $\overline{R}_{\alpha/2}$ and
$\overline{R}_{\alpha}$, as illustrated
in Fig.~\ref{fig1}.
\\

The first reflection,  $\overline{R}_{\alpha/2}$,
is through the plane containing the $K_3$ axis
and the line of the angle $\alpha/2$ in the
$K_1$--$K_2$ plane.
The second reflection, $\overline{R}_{\alpha}$,
is through the plane containing the $K_3$ axis and
the line of the angle $\alpha$ itself.
The arbitrary vector~$\tvec{K}$ goes into
$\widetilde{\tvec{K}}$ under the reflection
$\overline{R}_{\alpha/2}$ and further into
$\tvec{K}'$ under~$\overline{R}_{\alpha}$.
In Fig.~\ref{fig1} we show the projections
of these vectors onto the $K_1$--$K_2$ plane.
Since the third component of the vectors are
not changed under~$\overline{R}_{\alpha/2}$ and
$\overline{R}_{\alpha}$ we see that~$\tvec{K}'$
gives exactly the vector rotated by
the angle~$\alpha$, that is,
$\tvec{K}'= R_\alpha\; \tvec{K}$, where $R_\alpha$
is given by \eq{eq37}.
To see all
this formally we consider~$R_\beta$ as in~\eq{eq37} but
with $\alpha$ replaced by an arbitrary angle $\beta$ and define
\begin{equation}
\label{eq39}
\overline{R}_\beta = R_\beta \overline{R}_2 R_\beta^\trans \,=\,
\begin{pmatrix} \cos2\beta & \quad\phm\sin2\beta & \quad 0 \\
\sin2\beta &\quad  -\cos2\beta & \quad 0 \\
 0 & \quad 0 & \quad 1 \end{pmatrix}.
\end{equation}
Note that $(\overline{R}_\beta)^2=\mathds{1}_3$, which indicates
that $\overline{R}_\beta$ is a pure reflection.  In particular,
$\overline{R}_\beta$ describes the reflection through the plane
containing the $K_3$ axis and the line of angle~$\beta$
in the $K_1$--$K_2$ plane.
We have~$R_\beta^\trans =R_{- \beta}$,
$\overline{R}_2 R_\beta \overline{R}_2 = R_{-\beta}$, and
$R_{\beta_1} R_{\beta_2} = R_{\beta_1 + \beta_2}$.
It is then easy to see that we get $R_\alpha$ from~\eq{eq37}
via two reflections,
\begin{equation}
\label{eqmain}
\overline{R}_\alpha \overline{R}_{\alpha/2} =
R_\alpha \overline{R}_2 R^\trans_\alpha
R_{\alpha/2} \overline{R}_2 R^\trans_{\alpha/2} = R_{\alpha}.
\end{equation}
This is the $K$ space equivalence of the statement that any
Higgs family transformation is equivalent
to the product of two CP1 transformations.

\vskip 0.1in

\textbf{Proof of Theorem~2}.
If we perform three successive CP1 transformations
of the form given by \eq{eq27}, with $X$ given by $X_1$, $X_2$ and
$X_3$, respectively, then the resulting transformation is of the form
of \eq{eq27} with
\beq \label{xxx}
X=X_3 X_2^* X_1\,,
\eeq
where the $X_i$ are symmetric unitary matrices. However, any
unitary matrix $X$ can be written in the form of \eq{xxx}.  This
follows from the fact that any unitary matrix can be written as the
product of two symmetric unitary matrices, as proven in \eq{ss}.
Thus, one can simply choose one of the matrices $X_i$ in \eq{xxx} to
be the identity matrix\footnote{Of course, this does \textit{not}
imply that a GCP transformation can be expressed as the product of
two CP1 transformations.  Each CP1 transformation involves the complex
conjugation of the scalar fields, so one requires a product of odd number of
CP1 transformations to express a GCP transformation.  The product of
two CP1 transformations is a Higgs family transformation as demonstrated in
Theorem~1.}
(which is of course a symmetric unitary matrix).
Finally, we note that the product of an odd number of GCP
transformations is a GCP transformation, as this follows trivially
from \eq{eq27}.  The proof of Theorem 2 is then complete.

Again, it is instructive to prove Theorem 2 in $K$ space.  Consider an
arbitrary GCP transformation specified by \eq{eq29}.
In a suitable basis~$\overline{R}$ has the form in~\eq{eq35}.
For $\theta=0$ we are finished, since $\overline{R}=\overline{R}_2$ which
corresponds to a CP1 transformation.
For $\theta \neq 0$ we make a basis transformation
in~\eq{eq35} exchanging the 2 and the 3 axes, which yields an improper
rotation,
\begin{equation}
\label{eq26}
\overline{R}^{\,\prime}_{2\theta}=
\begin{pmatrix}
\cos 2 \theta & \quad -\sin 2 \theta & \quad \phm 0\\
\sin 2 \theta & \quad \phm\cos 2 \theta & \quad \phm 0 \\
0 & \quad 0 & \quad -1
\end{pmatrix}.
\end{equation}
We can represent this matrix as a product
of three CP1 transformations by multiplying
\eq{eqmain} by $\overline{R}_3$ and replacing
$\alpha$ by $2 \theta$. Thus we obtain
\begin{equation}
\label{eq42}
\overline{R}_3 \overline{R}_{2 \theta} \overline{R}_{\theta}
= \big( \widetilde{R} \overline{R}_2 \widetilde{R}^\trans \big)
\big( R_{2 \theta} \overline{R}_2 R^\trans_{2 \theta} \big)
\big( R_\theta \overline{R}_2 R^\trans_\theta \big)
= \overline{R}^{\,\prime}_{2\theta}\,,
\end{equation}
where
\begin{equation}
\widetilde{R}=
\begin{pmatrix}
1 & \,\,\,\phm 0 &\,\,\,\phm  0\\
0 & \,\,\, \phm 0 &\,\,\, -1\\
0 & \,\,\,\phm 1 &\,\,\,\phm  0
\end{pmatrix}.
\end{equation}
\Eq{eq42} is the $K$ space equivalence of the statement that an arbitrary GCP transformation
either is a CP1 transformation or
the product of three CP1 transformations.

Note that Theorem 2 is equivalent to the statement that an arbitrary improper $3\times 3$ rotation
matrix can be expressed as a product of three reflection matrices.
This statement can be proved directly.
First, we note that any proper $3\times 3$ rotation matrix can be written as the product of
two reflection matrices as in \eq{eqmain}.  However, any improper rotation matrix can be written
as the product of a reflection matrix and a proper rotation matrix.  Combining these two statements
yields the desired result.\footnote{In particular,
the inversion matrix $-\mathds{1}_3$ can be written as the
product of three reflection matrices.  Starting from \eq{eqmain}, we note that $-R_\alpha^\trans$ is
a reflection matrix, in which case,
$-R_\alpha^\trans \overline{R}_\alpha\overline{R}_{\alpha/2}=-\mathds{1}_3$.}

\section{Higgs-family and GCP symmetries}

In Sections 3 and 4, we analyzed Higgs-family and GCP {\em transformations}.
In this section, we shall address the corresponding {\em symmetries}.
In the scalar sector of the THDM, there are six nontrivial inequivalent classes
of symmetries that can be exhibited by the tree-level THDM scalar potential.
The complete classification of the THDM symmetry classes, which are listed in Table I,
was first given in \cite{Ivanov:2007de} and subsequently analyzed in \cite{Ferreira:2009wh}.

The following points are noteworthy.
First, consider the case in which a Higgs-family or a GCP transformation
is a symmetry transformation of the THDM. The representation
of such a symmetry transformation as a product of other transformations,
(e.g. a product of CP1 transformations as discussed in Section 4),
does {\em not} automatically
imply that the individual factors of the product are also
symmetries of the THDM.
Second, the analysis of this paper is confined to the
scalar sector of the THDM.  Ultimately, one must also
include the Yukawa couplings to fermions in the theory in the
discussion of symmetry transformations.
The structure of the symmetries of the scalar sector may not be
respected by the Yukawa sector.  As an example,
consider a field transformation $S_1$ that can be written as a
product of two others, $S_1 = S_2 S_3$. Suppose that by imposing
the symmetry $S_1$, the symmetries $S_2$ and $S_3$
are automatically respected by the scalar sector.
This property is not guaranteed to
hold for the Yukawa sector. In particular, it is possible that
imposing $S_2$ and $S_3$ separately
as symmetries may lead to stronger restrictions as
compared with the imposition of the symmetry $S_1$ by itself.
This is indeed the case in the model
which was studied in~\cite{Maniatis:2007de,Maniatis:2009vp,Maniatis:2009by}.

\begin{table}[t!]
\begin{tabular}{lll}
\hline
symmetry class \qquad\quad & rotation matrices in a generic basis \qquad\quad & rotation matrices as a product of CP1 reflection matrices\\
\hline
$\mathbb{Z}_2$ &$\phm R(\boldsymbol{\hat n},\pi)$ & $R_\pi= \overline{R}_2 \overline{R}_1$, for $\boldsymbol{\hat n}=\boldsymbol{\hat z}$ [see~\eq{eq31n}]\\
U(1) &$\phm R(\boldsymbol{\hat n}, \theta)\,,\quad (0<\theta<\pi)$ & $R_{2 \theta}= \overline{R}_{2 \theta} \overline{R}_{\theta}$,
for $\boldsymbol{\hat n}=\boldsymbol{\hat z}$ [see~\eq{eq36}]\\
SO(3) &$\pm R(\boldsymbol{\hat n}_1, \theta)\,,\, \pm
R(\boldsymbol{\hat n}_2, \theta)$\,,\,
($\boldsymbol{\hat n}_1\boldsymbol{\times}\boldsymbol{\hat n}_2\neq 0$)&
See section \ref{sec:diag} \\
CP1 & $-R(\boldsymbol{\hat n},\pi)$ & $\overline{R}_2$, for $\boldsymbol{\hat n}=\boldsymbol{\hat y}$
[see \eq{cp1y}]\\
CP2 & $-R(\boldsymbol{\hat n},0)=-\unitmatrix_3$  & $-\unitmatrix_3=\overline{R}_3 \overline{R}_2 \overline{R}_1$ [see \eq{eq42n}]\\
CP3 & $-R(\boldsymbol{\hat n}, \theta)\,,\quad(0<\theta<\pi)$ & $\overline{R}'_{2\theta_1}=\overline{R}_3 \overline{R}_{2\theta_1} \overline{R}_{\theta_1}$,
for $\boldsymbol{\hat n}=\boldsymbol{\hat z}$ [see \eq{eq:rcp3}] \\
\hline
\end{tabular}
\caption{\label{tab1} \small
The symmetry classes and the corresponding
proper [and improper] rotation matrices $R$
[and $\overline{R}$] that generate the symmetry classes in $K$ space.
A general rotation matrix $R(\axis,\theta)$ is uniquely determined by an
axis of rotation $\axis$ and rotation angle $\theta$ (see Appendix A).
The identity class generated by $R(\boldsymbol{\hat n},0)=\unitmatrix_3$ is trivial
and is not explicitly displayed.
The decomposition of $R$ and $\overline{R}$
in terms of products of CP1 reflection matrices is also
given in a particular basis.
In the case of the CP2 and SO(3) symmetry classes,
the corresponding rotation matrices are invariant with respect to
basis transformations.  The generation of the SO(3) symmetry class
requires two rotation matrices (each of which may be a proper or
improper rotation depending on the overall sign choice), where
$\boldsymbol{\hat n}_1$ and $\boldsymbol{\hat n}_2$ are non-collinear.}
\end{table}

\subsection{Symmetries and single rotations in bilinear space}

We shall now use Theorems 1 and 2 of Section 4 to obtain a new
derivation of the statement (given in~\cite{Ferreira:2009wh})
that all possible symmetries of the scalar sector
of the THDM can be reduced to multiple applications
of the standard CP symmetry in suitable bases.
In Table~\ref{tab1} we list the possible classes of THDM symmetries,
along with the equivalent result in $K$ space, where the
Higgs family and generalized CP transformations are generated by
proper or improper rotations, $R$ or $\overline{R}$, respectively.
Also shown is the decomposition
of $R$ and $\overline{R}$ in terms of products of CP1 reflection matrices
in a particular basis choice.\footnote{In
the CP2 symmetry class, the rotation matrices are independent of $\boldsymbol{\hat n}$, so
that the basis choice in this case is fixed by the direction of an eigenvector of $E$ corresponding
to one of its non-degenerate eigenvalues.}
Note that it is sufficient to
require the  invariance under a {\em single} but
suitable rotation in order to generate each symmetry class, with
the exception of the SO(3) case.\footnote{For the SO(3) symmetry
class, if a basis for the scalar fields is appropriately chosen,
then it is possible to generate the SO(3) symmetry class via
a single rotation matrix, as shown in Section 5.3.}

The transformation matrices $R$ and $\overline{R}$ generating the
respective symmetry classes can be
written as products of CP1 transformations using Theorems 1 and 2
of Section 4. Of course, one still needs to check if
the imposition of the CP1 factors as symmetries is
equivalent or more restrictive than the corresponding proper or
improper rotation, $R$ or
$\overline{R}$, alone.
Let us now derive and discuss the results shown in Table~\ref{tab1}
in detail.
\\

\begin{itemize}
\item \bf{$\mathbb{Z}_2$ symmetry}
\end{itemize}

The $\mathbb{Z}_2$ symmetry \cite{Glashow:1976nt,Paschos:1976ay},
\begin{equation}
\begin{aligned}
&\varphi_1(x) \to \phantom{+}\varphi_1(x),\\
&\varphi_2(x) \to -\varphi_2(x),
\end{aligned}
\label{zee2}
\end{equation}
corresponds in $K$ space to a rotation by $\pi$
around the third axis,
\begin{equation}
\label{eq30n}
\tvec{K}(x) \to R_\pi \tvec{K}(x)\,,
\end{equation}
where $R_\pi$ is as in~\eq{eq37}
with $\alpha=\pi$. That is,
\begin{equation}
R_{\mathbb{Z}_2} = R_\pi = \begin{pmatrix}
-1 & \phm 0 & \phm 0 \\
 \phm 0 & -1 & \phm 0 \\
 \phm 0 & \phm 0 & \phm 1
\end{pmatrix}.
\end{equation}
This result is most easily obtained by taking $U=\sigma^3$ in \eq{Rabu}.

The application of
Theorem 1 gives the decomposition [cf.~\eq{eqmain}]:
\begin{equation}
\label{eq31n}
R_\pi = \overline{R}_\pi \overline{R}_{\pi/2} = \overline{R}_2\overline{R}_1\,.
\end{equation}
Requiring $\mathbb{Z}_2$ to be a symmetry means that the THDM
parameters specified in \eq{eq17}
must satisfy~\eq{eq18} with $R(U)$
replaced by $R_\pi$, which yields:
\begin{equation}
\label{eq32n}
\xi_1=\xi_2=0 ,\quad
\eta_1=\eta_2=0,\quad
E_{13}=E_{23}=0 .
\end{equation}
On the other hand, requiring the CP1 transformations corresponding
to both $\overline{R}_1$ and $\overline{R}_2$ to be separate symmetries
gives, in addition to~\eq{eq32n},
\begin{equation}
E_{12}=0 .
\end{equation}
Thus, imposing the CP1 symmetries $\overline{R}_1$ and $\overline{R}_2$  yields
a \textit{stronger} constraint than the $\mathbb{Z}_2$ symmetry alone.
Nevertheless, the scalar potential subject to the
CP1 symmetries $\overline{R}_1$ and $\overline{R}_2$ is \textit{physically}
equivalent to the scalar potential subject to the $\mathbb{Z}_2$
symmetry, since the two scalar potentials are relate by a change of
basis.  This is easily proved by performing a change of basis
characterized by an SO(3) matrix $R$ [cf.~\eq{eq12}] with
$R_{13}=R_{23}=R_{31}=R_{32}=0$ and $R_{33}=1$.  Note that $\boldsymbol{\xi}$ and
$\boldsymbol{\eta}$ are invariant under this change of basis,
whereas $E\to RER^\trans$.  One can choose $R$ such that $E$ is diagonal,
which confirms that the two scalar potentials are related by a
basis transformation.

It is also instructive to introduce the permutation symmetry $\Pi_2$,
\begin{equation}
\begin{aligned}
&\varphi_1(x) \to \varphi_2(x),\\
&\varphi_2(x) \to \varphi_1(x).
\end{aligned}
\end{equation}
In fact, the $\mathbb{Z}_2$-symmetric scalar potential and
the $\Pi_2$-symmetric scalar potential are related by a basis
transformation~\cite{Davidson:2005cw,Ferreira:2009wh}
(and are hence physically equivalent).
To obtain the $K$ space description of $\Pi_2$, simply insert
$U=\sigma^1$ into \eq{Rabu}, which yields:
\begin{equation}
R_{\Pi_2} = \begin{pmatrix}
\phm 1 & \phm 0 & \phm 0 \\
 \phm 0 & -1 & \phm 0 \\
 \phm 0 & \phm 0 & -1
\end{pmatrix}.
\end{equation}
Requiring ${\Pi}_2$ to be a symmetry implies that the THDM
parameters specified in \eq{eq17}
must satisfy~\eq{eq18} with $R(U)$
replaced by $R_{\Pi_2}$, which yields:
\begin{equation}
\label{eq32n2}
\xi_2=\xi_3=0 ,\quad
\eta_2=\eta_3=0,\quad
E_{12}=E_{13}=0 .
\end{equation}

It is possible (although not particularly illuminating) to construct
the basis transformation that relates \eqs{eq32n}{eq32n2}.  However,
it is more useful to identify the most general transformation in $K$ space
that corresponds to the presence of the $\mathbb{Z}_2$ symmetry
specified by \eq{zee2} in some basis.  This can be accomplished by
starting in a basis where \eq{eq30n} is satisfied and transforming
to an arbitrary basis.

Under a basis transformation specified by the U(2) matrix $U$
[cf.~\eqs{eq11}{eq12}], we define $\widetilde{R}\equiv R(U)$,
where $R(U)$ is given by \eq{Rabu}.  Then, the $\mathbb{Z}_2$
symmetry transformation, $\tvec{K}(x) \to R_\pi \tvec{K}(x)$, is transformed to
\beq
\widetilde{R}\tvec{K}(x)\to\widetilde{R}R(\boldsymbol{\hat z},\pi)
  \widetilde{R}^\trans\widetilde{R}\tvec{K}(x)\,,
\eeq
where $R_\pi\equiv R(\boldsymbol{\hat z},\pi)$ is a rotation by 180$^\circ$
about the $z$-axis, and the expression $\widetilde{R}^\trans\widetilde{R}=\id$
has been conveniently inserted.  Using \eq{nz} given in Appendix A,
it follows that
\beq \label{Kp}
\tvec{K}^\prime(x)\equiv \widetilde{R}\tvec{K}(x)
= R(\boldsymbol{\hat n},\pi) \tvec{K}^\prime(x)\,,
\qquad \boldsymbol{\hat n}\equiv \widetilde{R}\boldsymbol{\hat z}\,,
\eeq
which is the form of the $\mathbb{Z}_2$ symmetry in the new basis.

Henceforth, we drop the primed superscripts.  The
most general transformation in $K$ space that corresponds to the
presence of the $\mathbb{Z}_2$ symmetry in some basis is given by:
\begin{equation}
\label{halfturn}
\tvec{K}(x) \to R(\boldsymbol{\hat n},\pi) \tvec{K}(x)\,,
\end{equation}
where $R(\boldsymbol{\hat n},\pi)$ is a rotation by $180^\circ$ about
an axis that is parallel to the unit
vector $\boldsymbol{\hat n}$.
As noted in Appendix A,
$R(\boldsymbol{\hat n},\pi)$ possesses one non-degenerate eigenvalue
equal to $+1$ and two degenerate eigenvalues $-1$.  The eigenvector
corresponding to the non-degenerate eigenvalue $+1$ is identified as the
rotation axis $\boldsymbol{\hat n}$, since
\beq \label{fixedaxis}
R(\boldsymbol{\hat n},\pi)\axis=\axis
\eeq
is just the geometrical statement that the rotation axis is unaffected
by the rotation.  Under
the symmetry governed by \eq{halfturn}, the THDM
parameters specified in \eq{eq17}
must satisfy~\eq{eq18} with $R(U)$
replaced by $R(\boldsymbol{\hat n},\pi)$, which yields:
\beqa
\text{$\boldsymbol{\xi}$ and $\boldsymbol{\eta}$ are eigenvectors
of $R(\boldsymbol{\hat n},\pi)$ with eigenvalue $+1$} \quad&
\Longleftrightarrow&\quad
\text{$\boldsymbol{\xi}$ and  $\boldsymbol{\eta}$ are parallel to
$\boldsymbol{\hat n}$}, \label{bc1}\\
ER(\boldsymbol{\hat n},\pi)=R(\boldsymbol{\hat n},\pi)E\quad &
\Longleftrightarrow&\quad
\text{$E\boldsymbol{\hat n}$ is parallel to $\boldsymbol{\hat n}$}\,.
\label{bc2}
\eeqa
\Eq{bc1} is a consequence of the fact the eigenvalue $+1$ of
$R(\boldsymbol{\hat n},\pi)$ is non-degenerate, which implies that
any vector $\boldsymbol{v}$ that satisfies $R(\boldsymbol{\hat n},\pi)
\boldsymbol{v}=\boldsymbol{v}$ must be proportional to $\boldsymbol{\hat n}$.
To derive \eq{bc2}, we note that $\boldsymbol{\hat n}$ is a simultaneous
eigenvector of $R(\boldsymbol{\hat n},\pi)$ and $E$.
In particular, $R(\boldsymbol{\hat n},\pi)E\boldsymbol{\hat n}=
ER(\boldsymbol{\hat n},\pi)\boldsymbol{\hat n}=E\boldsymbol{\hat n}$, where
the last step follows from \eq{fixedaxis}.  Hence either
$E\boldsymbol{\hat n}=0$ or $E\boldsymbol{\hat n}$ is
an eigenvector of $R(\boldsymbol{\hat n},\pi)$ with eigenvalue $+1$.
Since the latter is non-degenerate, it follows that\footnote{One can derive
\eq{eijn} more explicitly by employing \eq{hturn} in $ER(\boldsymbol{\hat n},\pi)=R(\boldsymbol{\hat n},\pi)E$.
It follows that $n_i n_j E_{jk}=n_k n_j E_{ji}$.   Using the fact that $E^\trans=E$, one obtains
$E_{ij}n_j=Cn_i$, where the constant of proportionality is identified as $C\equiv n_j E_{jk} n_k$. \label{fn}}
\beq \label{eijn}
E_{ij}n_j\propto n_i.
\eeq
One can easily check that \eqs{bc1}{eijn} reduce to \eq{eq32n} or \eq{eq32n2}
when $\boldsymbol{\hat n}=\boldsymbol{\hat z}$ or $\boldsymbol{\hat x}$,
respectively.

\Eq{bc2} implies that the eigenvectors of $E$ can be chosen to be simultaneous
eigenvectors of $R(\boldsymbol{\hat n},\pi)$.
Since $E$ is a real symmetric matrix, these eigenvectors can be chosen
to be orthonormal.  We denote these eigenvectors by
$\{\boldsymbol{\hat n}\,,\,\boldsymbol{\hat m}\,,\,
\boldsymbol{\hat n\times\hat m}\}$.  We have already noted that
$\boldsymbol{\hat n}$ is an eigenvector of $E$ by virtue of \eq{eijn}.
Thus, the other two eigenvectors of $E$ must satisfy:\footnote{In
general, the eigenvectors $\boldsymbol{\hat m}_1$ and
$\boldsymbol{\hat m}_2$ defined in \eq{hatm} are not expected to be
eigenvectors of $E$.  In this case, $\boldsymbol{\hat m}$ is some
linear combination of  $\boldsymbol{\hat m}_1$ and
$\boldsymbol{\hat m}_2$, and similarly for $\boldsymbol{\hat
  n\times\hat m}$. In particular, $E$ is not generally diagonal
with respect to the basis $\{\axis\,,\,\boldsymbol{\hat m}_1\,,\,
\boldsymbol{\hat m}_2\}$.\label{fner}}
\beq \label{eigenm}
R(\boldsymbol{\hat n},\pi)\boldsymbol{\hat m}
=-\boldsymbol{\hat m}\,,\qquad\qquad
R(\boldsymbol{\hat n},\pi)(\boldsymbol{\hat n\times\hat m})=
-\boldsymbol{\hat n\times \hat m}\,.
\eeq
Because of the two-fold degeneracy of the eigenvalue $-1$ of
$R(\boldsymbol{\hat n},\pi)$, it is possible to perform orthogonal
transformations within the two-dimensional subspace spanned by
$\boldsymbol{\hat m}$ and $\boldsymbol{\hat n\times\hat m}$ that
leave the form of the $\mathbb{Z}_2$ symmetry transformation given
in \eq{halfturn} unchanged.  This simply means that the
form of the $\mathbb{Z}_2$ symmetry transformation does not uniquely
specify the basis in $K$ space.  To fix the basis completely, one must specify
$\boldsymbol{\hat n}$ and the eigenvectors of $R(\axis,\pi)$ corresponding to
the two-fold degenerate eigenvalue $-1$.

In summary, the $\mathbb{Z}_2$ symmetry corresponds to
$\tvec{K}(x) \to R(\boldsymbol{\hat n},\pi) \tvec{K}(x)$ for some
choice of $\boldsymbol{\hat n}$.  Imposing this symmetry on the scalar
potential requires that
\beqa
&& \text{$\boldsymbol{\xi}$ and $\boldsymbol{\eta}$ are parallel to
$\boldsymbol{\hat n}$}\,,\\
&& \text{$E$ is diagonal  with respect to the basis
$\{\boldsymbol{\hat n}\,,\,\boldsymbol{\hat m}\,,\,
\boldsymbol{\hat n\times\hat m}\}$}\,,\label{Ediagm}
\eeqa
where $\boldsymbol{\hat m}$ is a simultaneous eigenvector of
$R(\axis,\pi)$ and $E$, with $R(\axis,\pi)\boldsymbol{\hat m}=
-\boldsymbol{\hat m}$.  In this case, the choice
of $\boldsymbol{\hat n}$ and $\boldsymbol{\hat m}$ uniquely fixes the
basis in $K$ space.
\\

\begin{itemize}
\item \bf{U(1) Peccei-Quinn symmetry}
\end{itemize}

The U(1) Peccei--Quinn symmetry~\cite{Fayet:1974pd,Peccei:1977hh,Peccei:1977ur}
requires invariance under
\begin{equation}
\begin{aligned}
\label{eq34}
& \varphi_1(x) \to e^{- i \theta} \varphi_1(x), \qquad  \qquad \qquad \qquad \\
&\varphi_2(x) \to e^{i \theta} \varphi_2(x)\;,
\end{aligned}
\end{equation}
with an arbitrary angle $\theta$, which is defined modulo $\pi$.
By taking
$U=\mathds{1}_2\cos\theta-i\sigma^3\sin\theta$ in \eq{Rabu}, we obtain
the corresponding symmetry transformation in $K$ space,
\begin{equation}
\label{Rpq}
\tvec{K}(x) \to R_{2\theta} \tvec{K}(x)\,,
\end{equation}
where $R_{2\theta}$ is as in~\eq{eq37}
with $\alpha=2\theta$, with $0\leq\theta<\pi$. That is,
\begin{equation}
R_{\rm U(1)} = R_{2\theta} = \begin{pmatrix}
 \cos 2\theta & -\sin 2\theta & \phm 0 \\
 \sin 2\theta & \phm\cos 2\theta & \phm 0 \\
 0 & \phm 0 & \phm 1
\end{pmatrix}\,.
\end{equation}

The application of
Theorem 1 gives the decomposition [cf.~\eq{eqmain}]:
\begin{equation}
\label{eq36}
R_{2\theta}=\overline{R}_{2\theta} \overline{R}_{\theta} .
\end{equation}
Requiring the U(1) Peccei-Quinn transformation to be a symmetry
implies that the THDM parameters specified in \eq{eq17}
must satisfy~\eq{eq18} with $R(U)$
replaced by $R_{2\theta}$ (for all possible values of $\theta$), which yields:
\begin{equation}
\label{eq35n}
\xi_1=\xi_2=0 ,\quad
\eta_1=\eta_2=0,\quad
E=\diag(\mu_1, \mu_1, \mu_3).
\end{equation}
It is straightforward to check that imposing the two CP1 symmetries
$\overline{R}_{\theta}$ and $\overline{R}_{2\theta}$ separately is equivalent
to requiring invariance under $R_{2\theta}$.

Remarkably, it is sufficient to require invariance
of the scalar potential given by \eq{eq5} under $R_{2\theta_0}$
for \textit{any} single particular value of
$2\theta_0\neq 0$ (mod~$\pi$)~\cite{Ferreira:2008zy}.\footnote{Note that if $\theta_0=\pi/2$
then $R_{2\theta_0}=R_\pi$, which
generates the $\mathbb{Z}_2$ symmetry class treated previously.
What is perhaps more surprising is that if $\theta_0=\pi/n$ for any
integer $n>2$, then the invariance of the scalar potential
under $R_{2\theta_0}$ [which generates a $\mathbb{Z}_n$
subgroup of U(1)] implies invariance under the full U(1)
group.  This latter result is a consequence of the fact that
the scalar potential of \eq{eq5} contains no terms of dimension
greater than four.  If one relaxes this condition, then new symmetry
classes can arise that are associated with discrete subgroups of U(1)
of order $d>2$.}
That is, invariance under $R_{2\theta_0}$ for \textit{any}
$2\theta_0\neq 0$ (mod~$\pi$) implies \eq{eq35n}, which in
turn implies invariance under $R_{2\theta}$ for \textit{all}
values of $\theta$.

More generally, we consider the possibility that a basis transformation
is required to identify the symmetry specified in \eq{eq34}.
First, we shall rename $2\theta$ by $\theta$ and rewrite \eq{Rpq}
as
\beq \label{Rpq2}
\tvec{K}(x) \to R(\pm\boldsymbol{\hat z},\theta) \tvec{K}(x)\,,
\eeq
where $0\leq\theta\leq \pi$.  Note that we must allow for both signs of
$\pm\boldsymbol{\hat z}$ in order to cover the entire U(1) Peccei-Quinn group manifold.
Under a basis transformation specified by the U(2) matrix $U$
[cf.~\eqs{eq11}{eq12}], we define $\widetilde{R}\equiv R(U)$,
where $R(U)$ is given by \eq{Rabu}.  Then, with assistance from \eq{nz}, the U(1) Peccei-Quinn
symmetry transformation, \eq{Rpq2}, is transformed to
\beq
\tvec{K}^\prime(x) \to \widetilde{R}R(\pm\boldsymbol{\hat z},\theta)
\widetilde{R}^\trans\tvec{K}^\prime(x)=R(\pm\boldsymbol{\hat n},\theta)\tvec{K}^\prime(x)\,,
\eeq
where $\axis=\widetilde{R}\boldsymbol{\hat z}$
and $R(\boldsymbol{\hat n},\theta)$ is a rotation by $\theta$ about
the axis $\boldsymbol{\hat n}$.

Thus, dropping the primed subscripts,
the most general transformation in $K$ space
that corresponds to the presence of the U(1) Peccei-Quinn symmetry in
some basis is given by:
\begin{equation}
\label{rnt}
\tvec{K}(x) \to R(\boldsymbol{\hat n},\theta) \tvec{K}(x)\,,
\qquad \text{for}~ 0<\theta<\pi\,.
\end{equation}
Note that we have excluded the case of
$\theta=0$, which corresponds to the identity transformation,
and the case of
$\theta=\pi$, which corresponds to the $\mathbb{Z}_2$ symmetry
transformation treated previously.  When $\theta\neq 0~({\rm mod}~\pi)$,
$R(\boldsymbol{\hat n},\theta)$
possesses three non-degenerate eigenvalues:
$+1$, $e^{i\theta}$ and $e^{-i\theta}$, and $\boldsymbol{\hat n}$ is the
normalized eigenvector of $R(\boldsymbol{\hat n},\theta)$ with
eigenvalue $+1$.  It is convenient to introduce normalized eigenvectors
$\boldsymbol{\hat m}$ and $\boldsymbol{\hat m}^*$,
corresponding to the eigenvalues $e^{i\theta}$ and $e^{-i\theta}$, respectively.
For further details, see Appendix A.

Under the symmetry governed by \eq{rnt}, the THDM parameters specified in \eq{eq17}
must satisfy~\eq{eq18} with $R(U)$ replaced by $R(\boldsymbol{\hat n},\theta)$,
which yields:
\beqa
\hspace{-0.3in}\text{$\boldsymbol{\xi}$ and $\boldsymbol{\eta}$ are eigenvectors
of $R(\boldsymbol{\hat n},\theta)$ with eigenvalue $+1$} \quad&
\Longleftrightarrow&\quad
\text{$\boldsymbol{\xi}$ and  $\boldsymbol{\eta}$ are parallel to
$\boldsymbol{\hat n}$}, \label{pq1}\\
ER(\boldsymbol{\hat n},\theta)=R(\boldsymbol{\hat n},\theta)E\quad &
\Longleftrightarrow&\quad
\text{$E\boldsymbol{\hat v}$ is parallel to $\boldsymbol{\hat v}$
for $\boldsymbol{\hat v}=\boldsymbol{\hat n}\,,\,\boldsymbol{\hat m}\,,\,\boldsymbol{\hat m}^*$.}
\label{pq2}
\eeqa

The derivation of \eq{pq2} is similar to the one given in
the case of the $\mathbb{Z}_2$ symmetry above.  We thereby obtain two
conditions on the matrix $E$.  The first condition,
\beq \label{eipq}
E_{ij}n_j\propto n_i\,,
\eeq
coincides with \eq{eijn}.
To derive the second condition, we note that
$R(\boldsymbol{\hat n},\theta)E\boldsymbol{\hat m}=
ER(\boldsymbol{\hat n},\theta)\boldsymbol{\hat
  m}=e^{i\theta}E\boldsymbol{\hat m}$, which implies that either
$E\boldsymbol{\hat m}=0$ or $E\boldsymbol{\hat m}$ is an eigenvector
of $R(\boldsymbol{\hat n},\theta)$ with eigenvalue $e^{i\theta}$.
Since $\theta\neq 0$ (mod~$\pi$) by assumption, the latter is
non-degenerate, and it follows that\footnote{One
can derive \eq{eiipq} more explicitly by employing \eq{Rij} in
$ER(\boldsymbol{\hat n},\theta)=R(\boldsymbol{\hat n},\theta)E$.
This yields two constrains: $n_i n_j E_{jk}=n_k n_j E_{ji}$ (also
obtained in the case of the $\mathbb{Z}_2$ symmetry, as noted in
footnote \ref{fn}) and $\epsilon_{ij\ell}E_{jk}
n_\ell=\epsilon_{jk\ell}E_{ij}n_\ell$.  If we multiply the latter
equation by $m_k$ and employ $E^\trans=E$, $E\boldsymbol{\hat
n}\propto\boldsymbol{\hat n}$ and $\boldsymbol{\hat n}\boldsymbol{\cdot\hat
m}=0$, it then follows that $E_{ij}m_j=\kappa m_i$, where the constant of
proportionality is identified as $\kappa \equiv \tfrac{1}{2}({\rm
Tr}~E-n_iE_{ij}n_j)$.
 }
\beq \label{eiipq}
E_{ij}m_j\propto m_i.
\eeq
This equation and its complex conjugate imply that $\boldsymbol{\hat
  m}$ and $\boldsymbol{\hat m}^*$ are eigenvectors of $E$, whose
eigenvalues are complex conjugates of each other.  But $E$ is a real
symmetric matrix, which implies that its eigenvalues are real.  It
follows that there is (at least) a two-fold degeneracy among the
eigenvalues of $E$.

As noted below \eq{m2}, the eigenvector $\boldsymbol{\hat m}$ is
\textit{independent} of the rotation angle $\theta$, assuming that
$\theta\neq 0$ (mod~$\pi$).   Consequently, the constraints on the
scalar potential parameters [governed by eqs.~(\ref{pq1}), (\ref{eipq})
and (\ref{eiipq})] do not depend on $\theta$.  That is,
the invariance of the scalar potential under
$R(\boldsymbol{\hat n},\theta)$ for \textit{any} single particular value of $\theta\neq 0$
(mod~$\pi$) yields the U(1)
Peccei-Quinn symmetry, which in turn implies the invariance
of the scalar potential under
$R(\boldsymbol{\hat n},\theta)$ for \textit{all} values of $\theta$.

One can check that
eqs.~(\ref{pq1}), (\ref{eipq}) and (\ref{eiipq}) reduce to
\eq{eq35n} in the basis where
$\boldsymbol{\hat n}=\boldsymbol{\hat z}$ and
$\boldsymbol{\hat m}=\tfrac{1}{\sqrt{2}}(\boldsymbol{\hat
  x}-i\boldsymbol{\hat y})$.  For example, after inserting 
$\boldsymbol{\hat m}$ into \eq{eiipq}, and taking into
account that $E$ is a real symmetric matrix, we obtain:
\beq
E_{11}-iE_{12}=iE_{12}+E_{22}\,,\qquad E_{13}-iE_{23}=0\,.
\eeq
Taking the real and imaginary parts
of the above equations, it follows that
$E$ is a diagonal matrix with $E_{11}=E_{22}$ as stated in \eq{eq35n}.
Indeed, the above computation is valid for any choice of $\theta\neq 0$
(mod~$\pi$), as noted above.
Due to \eqs{eipq}{eiipq}, the vectors comprising the
orthonormal set $\{\boldsymbol{\hat n}\,,\boldsymbol{\hat m}\,,
\boldsymbol{\hat m}^*\}$ are simultaneous eigenvectors of $R(\axis,\theta)$ and $E$.
This means that $E$ is diagonal with respect to the basis $\{\boldsymbol{\hat n}\,,
\boldsymbol{\hat m}\,,\boldsymbol{\hat m}^*\}$.  Since $\boldsymbol{\hat m}$
and $\boldsymbol{\hat m}^*$ are uniquely determined by $\axis$ [cf.~\eqs{m1}{m2}]
when $\theta\neq 0$ (mod~$\pi$), it follows that $\axis$ uniquely fixes the basis
in $K$ space.
\\

\begin{itemize}
\item \bf{CP1 symmetry}
\end{itemize}

In Section 3, we showed that one can always find a basis in which the
CP1 transformation in $K$ space is given by
\begin{equation}
\label{RCP1}
\tvec{K}(x) \to \overline{R}_2 \tvec{K}(x')\,,
\end{equation}
That is, 
\begin{equation} \label{cp1y}
\overline{R}_{\rm CP1} = \overline{R}_2 = \begin{pmatrix}
1 & \phm 0 & \phm 0 \\
 0 & -1 & \phm 0 \\
 0 & \phm 0 & \phm 1
\end{pmatrix}.
\end{equation}
Requiring CP1 to be a symmetry implies that the THDM
parameters specified in \eq{eq17}
must satisfy~\eq{GCPsymm} with $\overline{R}\equiv\overline{R}_{\rm CP1}$, 
which yields:
\begin{equation} \label{cp1cond}
\xi_2 = 0\; , \; \eta_2 = 0 \; , \; E_{12} = E_{23} = 0.
\end{equation}
As expected, \eq{cp1cond} is equivalent to the statement that
in a basis in which the CP1 transformation is given by \eq{RCP1},
all the parameters of the scalar potential specified in \eq{eq5}
are real.

As in the case of the $\mathbb{Z}_2$ symmetry, there still remains
some freedom to perform a basis transformation while maintaining the
form of the CP1 symmetry given in \eq{RCP1}.  By performing a change
of basis characterized by an SO(3) matrix $R$ with
$R_{12}=R_{23}=R_{21}=R_{32}=0$ and $R_{22}=1$, we see that \eq{cp1cond} is still
satisfied, while other matrix elements of $E$ are transformed
according to $E\to RER^\trans$.  One is free to choose
$R$ such that $E$ in the new basis is diagonal.

The above results pertain to a specific basis choice.
With respect to an arbitrary basis, we showed in Section 3 that
$\overline{R}_{\rm CP1}$  is a reflection through planes in $K$ space,
which implies that $\overline{R}_{\rm CP1}$ is an improper rotation
matrix that satisfies
\beq \label{Rbar1}
\overline{R}_{\rm CP1}^2=\mathds{1}_3\,,\qquad \overline{R}_{\rm
  CP1}\neq -\mathds{1}_3\,.
\eeq
An explicit form is given by:
\beq
\overline{R}_{\rm CP1}=-R(\boldsymbol{\hat n},\pi)\,,
\eeq
where the unit vector $\boldsymbol{\hat n}$ points in the direction
normal to the reflection plane.  In general,
$-R(\boldsymbol{\hat n},\pi)$ possesses one non-degenerate eigenvalue $-1$
and a doubly-degenerate eigenvalue $+1$.   We identify
$\boldsymbol{\hat n}$ as the eigenvector of $-R(\boldsymbol{\hat n},\pi)$
corresponding to the non-degenerate eigenvector $-1$.  The eigenvectors
corresponding to the doubly-degenerate eigenvalues
of $\overline{R}_{\rm CP1}$ span the reflection
plane in $K$ space.

The scalar potential exhibits the CP1 symmetry, \eq{RCP1}, in some
basis if
\begin{equation}
\label{RCP1g}
\tvec{K}(x) \to -R(\boldsymbol{\hat n},\pi) \tvec{K}(x')
\end{equation}
is a symmetry. That is, the THDM
parameters specified in \eq{eq17}
must satisfy~\eq{GCPsymm} with $\overline{R}\equiv\overline{R}_{\rm CP1}$, 
which yields:
\beqa
\text{$\boldsymbol{\xi}$ and $\boldsymbol{\eta}$ are eigenvectors
of $-R(\boldsymbol{\hat n},\pi)$ with eigenvalue $+1$} \quad&
\Longleftrightarrow&\quad
\text{$\boldsymbol{\xi}$ and  $\boldsymbol{\eta}$ are perpendicular to
$\boldsymbol{\hat n}$}, \label{cp11}\\
ER(\boldsymbol{\hat n},\pi)=R(\boldsymbol{\hat n},\pi)E\quad &
\Longleftrightarrow&\quad
\text{$E\boldsymbol{\hat n}$ is parallel to $\boldsymbol{\hat n}$}\,.
\label{cp12}
\eeqa
Following the derivation of \eq{eijn},
either $E\boldsymbol{\hat n}=0$ or $E\boldsymbol{\hat n}$ is an eigenvector
of $-R(\boldsymbol{\hat n},\pi)$ with eigenvalue $-1$.  Since the latter is non-degenerate,
we again recover \eq{eijn}.
It is a simple matter to confirm
that \eqs{eijn}{cp11} reduce to \eq{cp1cond} when $\boldsymbol{\hat n}=
\boldsymbol{\hat y}$.

We can introduce
simultaneous eigenvectors of $E$ and $-R(\axis,\pi)$, denoted by
$\boldsymbol{\hat m}$ and $\boldsymbol{\hat n\times\hat m}$,
which satisfy \eq{eigenm} [cf.~footnote~\ref{fner}].
These two vectors correspond to the two-fold degenerate
eigenvalue $+1$ of $-R(\axis,\pi)$.
As in the treatment of the $\mathbb{Z}_2$ symmetry,
the form of the CP1 symmetry transformation does not uniquely
specify the basis in $K$ space.  To fix the basis completely, one must specify
$\boldsymbol{\hat n}$ and the eigenvectors of $-R(\axis,\pi)$ corresponding to
the two-fold degenerate eigenvalue $+1$.

In summary, the CP1 symmetry corresponds to
$\tvec{K}(x) \to -R(\boldsymbol{\hat n},\pi) \tvec{K}(x')$ for some
choice of $\boldsymbol{\hat n}$.  Imposing this symmetry on the scalar
potential requires that
\beqa
&& \text{$\boldsymbol{\xi}$ and $\boldsymbol{\eta}$ are perpendicular to
$\boldsymbol{\hat n}$}\,,\label{perpto} \\
&& \text{$E$ is diagonal  with respect to the basis
$\{\boldsymbol{\hat n}\,,\,\boldsymbol{\hat m}\,,\,
\boldsymbol{\hat n\times\hat m}\}$}\,,
\eeqa
where $\boldsymbol{\hat m}$ is a simultaneous eigenvector of
$-R(\axis,\pi)$ and $E$, with $-R(\axis,\pi)\boldsymbol{\hat m}=
\boldsymbol{\hat m}$.  In this case, the choice
of $\boldsymbol{\hat n}$ and $\boldsymbol{\hat m}$ uniquely fixes the
basis in $K$ space.
\\

\begin{itemize}
\item \bf{CP2 symmetry}
\end{itemize}

The CP2 symmetry class is generated by requiring
that the scalar potential parameters specified in \eq{eq17} are invariant
under $\overline{R}=-R(\boldsymbol{\hat n},0)=-\unitmatrix_3$ corresponding
to $2\theta= \pi$ in~\eq{eq26}, which yields
\begin{equation}
\overline{R}_{\rm CP2} =
\overline{R}'_\pi=
\begin{pmatrix}
-1 & \phm 0 & \phm 0\\
\phm 0 & -1 & \phm 0 \\
\phm 0 &\phm 0 & -1
\end{pmatrix}.
\end{equation}
The form of $\overline{R}_{\rm CP2}$ is basis-independent, since
$R(U)\overline{R}_{\rm CP2}R^\trans(U)=\overline{R}_{\rm CP2}$ for any choice of $U$.
We may now invoke Theorem 2 from Section 4.
From~\eq{eq42}, we then obtain
\begin{equation}
\label{eq42n}
\overline{R}_3 \overline{R}_2 \overline{R}_1 = -\unitmatrix_3 .
\end{equation}
Invariance under $\overline{R}_{\rm CP2}=-\unitmatrix_3$ requires
$\tvec{\xi}=0$ and $\tvec{\eta}=0$ but leaves
$E$ arbitrary. We can then perform a basis transformation to diagonalize
$E$, which leaves $\tvec{\xi}$ and $\tvec{\eta}$
unchanged.  Invariance under all three individual CP1
transformations $\overline{R}_3$, $\overline{R}_2$, and $\overline{R}_1$,
taken together, implies that
\beq
\tvec{\xi}=\tvec{\eta}=0\quad {\rm and} \quad E=\diag(\mu_1,\mu_2,\mu_3)\,.
\eeq
Clearly,
$\overline{R}_3$, $\overline{R}_2$, and $\overline{R}_1$, when separately applied, yield
a {\em stronger} constraint than $-\unitmatrix_3$ alone.
But the three CP1 symmetries are physically {\em equivalent}
to $-\unitmatrix_3$, since the scalar potentials that result
from the two procedures are related by the change of basis that diagonalizes $E$.
As shown in \cite{Maniatis:2007de}, this equivalence may no longer hold
once the Yukawa sector is taken into account.\\

\begin{itemize}
\item \bf{CP3 symmetry}
\end{itemize}

It is convenient to choose a basis in which the CP3 symmetry
class is generated as follows. We require the transformation
$\overline{R}'_{2\theta_1}$ [defined in \eq{eq26}] to be a symmetry, for any
choice of angle such that $2\theta_1\neq 0$ (mod~$\pi$).
Theorem 2 can again be employed to express $\overline{R}'_{2\theta_1}$
as a product of three CP1 transformations.
\Eqs{eq26}{eq42} then yield:
\begin{equation}
\overline{R}_{\rm CP3} = \overline{R}'_{2\theta_1}=
\begin{pmatrix}
\cos 2 \theta_1 & -\sin 2 \theta_1 & \phm 0\\
\sin 2 \theta_1 & \phm \cos 2 \theta_1 & \phm 0 \\
0 & \phm 0 & -1
\end{pmatrix} = \overline{R}_3 \overline{R}_{2\theta_1}
\overline{R}_{\theta_1}.
\label{eq:rcp3}
\end{equation}

Requiring CP3 to be a symmetry implies that the THDM
parameters specified in \eq{eq17}
must satisfy~\eq{GCPsymm} with 
$\overline{R}\equiv\overline{R}^\prime_{2\theta_1}$, which yields:
\begin{equation} \label{cp3cond}
\boldsymbol{\xi}=\boldsymbol{\eta} = 0\; ,\qquad \;  E={\rm diag}(\mu_1\,,\,\mu_1\,,\,\mu_3)\,.
\end{equation}
It is straightforward to check that imposing the three CP1
symmetries $\overline{R}_3$, $\overline{R}_{2\theta_1}$ and
$\overline{R}_{\theta_1}$
separately is equivalent to requiring invariance under
$\overline{R}^\prime_{2\theta_1}$.
Similarly to the case of the U(1) Peccei-Quinn symmetry,
the invariance of the scalar potential under $\overline{R}^\prime_{2\theta_1}$,
for \textit{any} single particular value of
$2\theta_1\neq 0$ (mod~$\pi$),
implies invariance under $\overline{R}^\prime_{2\theta}$
for \textit{all} values of $\theta$.

The above results pertain to a specific basis choice.
With respect to an arbitrary basis, we showed in Section 3 that
$\overline{R}_{\rm CP3}$  is an improper rotation matrix that
satisfies $\overline R_{\rm CP3}^2\neq\mathds{1}_3$ (i.e., $\overline{R}_{\rm CP3}$
is \textit{not} a pure reflection in $K$ space).  The most general form
for the CP3 transformation is given by
$\overline{R}_{\rm CP3}=-R(\boldsymbol{\hat n},\theta)$ for $0<\theta<\pi$.
Thus the scalar potential, in a basis specified by $\boldsymbol{\hat n}$, exhibits the CP3 symmetry
if it is invariant under the transformation
\begin{equation}
\label{RCP3}
\tvec{K}(x) \to -R(\boldsymbol{\hat n},\theta)\tvec{K}(x')\,,\qquad \text{for}~~0<\theta<\pi\,.
\end{equation}
Note that we have excluded the case of $\theta=0$, which corresponds to the
CP2 symmetry transformation, and the case of $\theta=\pi$, which corresponds to
the CP1 symmetry transformation, both of which have already been treated above.
When $\theta\neq 0~({\rm mod}~\pi)$, the improper rotation matrix
$-\overline{R}(\boldsymbol{\hat n},\theta)$
possesses three non-degenerate eigenvalues:
$-1$, $-e^{i\theta}$ and $-e^{-i\theta}$, and $\boldsymbol{\hat n}$ is the
normalized eigenvector of $-R(\boldsymbol{\hat n},\theta)$ with
eigenvalue $-1$.   As in the analysis of the U(1) Peccei-Quinn symmetry,
it is convenient to introduce normalized eigenvectors of $-R(\boldsymbol{\hat n},\theta)$,
denoted by $\boldsymbol{\hat m}$ and $\boldsymbol{\hat m}^*$,
corresponding to the eigenvalues $-e^{i\theta}$ and $-e^{-i\theta}$, respectively.

Under the symmetry governed by \eq{RCP3}, the THDM parameters specified in \eq{eq17}
must satisfy~\eq{GCPsymm} with 
$\overline{R}\equiv -R(\boldsymbol{\hat n},\theta)$.  
Since $-R(\boldsymbol{\hat n},\theta)$
has no eigenvalue~$+1$ for $\theta\neq 0$ (mod~$\pi$), it follows that:
\beqa
\hspace{-0.3in} -R(\boldsymbol{\hat n},\theta) \tvec{\xi}= \tvec{\xi},\quad
-R(\boldsymbol{\hat n},\theta) \tvec{\eta}= \tvec{\eta}\;,
\quad& \Longleftrightarrow&\quad
\boldsymbol{\xi}=\boldsymbol{\eta}=0\,,
\label{cp31}\\
ER(\boldsymbol{\hat n},\theta)=R(\boldsymbol{\hat n},\theta)E\quad &
\Longleftrightarrow&\quad
\text{$E\boldsymbol{\hat v}$ is parallel to $\boldsymbol{\hat v}$
for $\boldsymbol{\hat v}=\boldsymbol{\hat n}\,,\,\boldsymbol{\hat m}\,,\,\boldsymbol{\hat m}^*$}.
\label{cp32}
\eeqa

The derivation of \eq{cp32} is identical to the one given in the analysis of the
U(1) Peccei-Quinn symmetry.  Thus, we again recover
the results of \eqs{eipq}{eiipq},
\beq \label{eicp3}
E_{ij}n_j\propto n_i\quad {\rm and} \quad E_{ij}m_j\propto m_i\,.
\eeq
For the same reasons given below \eq{eiipq},
there is (at least) a two-fold degeneracy among the
eigenvalues of $E$.
Since the constraints imposed by \eqs{cp31}{eicp3} are
\textit{independent} of the angle $\theta$,
the invariance of the scalar potential under
$-R(\boldsymbol{\hat n},\theta)$ for \textit{any}
value of $\theta\neq 0$
(mod~$\pi$) yields the CP3 symmetry,
which in turn implies the invariance of the scalar potential under
$-R(\boldsymbol{\hat n},\theta)$ for
\textit{all} values of $\theta$.
As expected, when  $\boldsymbol{\hat n}=
\boldsymbol{\hat z}$ and
$\boldsymbol{\hat m}=\tfrac{1}{\sqrt{2}}(\boldsymbol{\hat
  x}-i\boldsymbol{\hat y})$,
\eqs{cp31}{eicp3} lead to the previous results given in \eq{cp3cond}
for any value of $\theta\neq 0$ (mod~$\pi$).

By virtue of \eq{eicp3}, the vectors comprising
the orthonormal set $\{\boldsymbol{\hat n}\,,\boldsymbol{\hat m}\,,
\boldsymbol{\hat m}^*\}$ are simultaneous eigenvectors of $-R(\axis,\theta)$ and~$E$.
This means that $E$ is diagonal with respect to the basis $\{\boldsymbol{\hat n}\,,
\boldsymbol{\hat m}\,,\boldsymbol{\hat m}^*\}$.  Since $\boldsymbol{\hat m}$
and $\boldsymbol{\hat m}^*$ are uniquely determined by $\axis$ [cf.~\eqs{m1}{m2}]
when $\theta\neq 0$ (mod~$\pi$), it follows that $\axis$ uniquely fixes the basis
in $K$ space.

\begin{itemize}
\item \bf{The SO(3)-symmetric scalar potential}
\end{itemize}

Suppose that the THDM scalar potential is invariant under all possible
Higgs-family transformations [cf.~\eq{eq11}].  The U(2) invariance
group contains a gauged hypercharge U(1) subgroup, which is
automatically satisfied by the Higgs Lagrangian.  The ungauged
invariance group is therefore isomorphic to SO(3).  The elements of
this SO(3) group are in one-to-one correspondence with the matrices
$R(U)$ defined in \eq{Rabu}.  Hence, the invariance of the scalar
potential under SO(3) requires that \eq{eq18} must be satisfied for
\textit{any} SO(3) matrix $R(U)\in$~SO(3).  This clearly implies that
\begin{equation}
\label{eq37n}
\tvec{\xi}=0 ,\quad
\tvec{\eta}=0 ,\quad
E = \mu \unitmatrix_3\,,
\end{equation}
which of course must hold for any choice of basis (since the set of all basis transformations
has been promoted to a global symmetry of the scalar potential).
In $K$ space, one rotation matrix is not sufficient to generate all the
elements of the SO(3) symmetry class.
Since any rotation matrix must be of the form $R(\boldsymbol{\hat n},\theta)$, the
results of Section 5.1 imply that a single rotation matrix will generate either the
$\mathbb{Z}_2$ symmetry class (if $\theta=\pi$) or the U(1) Peccei-Quinn symmetry class,
if $0<\theta<\pi$.  However, one can generate all the elements of
the SO(3) symmetry class with two rotation matrices,
$R(\boldsymbol{\hat n}_1,\theta)$ and $R(\boldsymbol{\hat n}_2,\theta)$, where
$\boldsymbol{\hat n}_1\boldsymbol{\times}\boldsymbol{\hat n}_2\neq 0$.
In group theory language, this is
just the statement that starting from two non-trivial rotations for which
the corresponding rotation axes are non-collinear, it is
possible to generate \textit{any} element of SO(3) by successive multiplication of
the two initial rotation matrices and products thereof.

For example,
the SO(3) symmetry class can be generated through the application of both
\begin{equation}
R_{\alpha_1}=
\begin{pmatrix}
\cos \alpha_1 & -\sin \alpha_1 & \phm 0\\
\sin \alpha_1 & \phm\cos \alpha_1 & \phm 0\\
0 & \phm 0 & \phm 1\\
\end{pmatrix},
\qquad \text{with } 0 < \alpha_1 < \pi,
\end{equation}
and
\begin{equation}
R^\prime_{\alpha_2}=
\begin{pmatrix}
1 & \phm 0 & \phm 0 \\
0 & \phm\cos \alpha_2 & -\sin \alpha_2 \\
0 & \phm\sin \alpha_2 & \phm\cos \alpha_2
\end{pmatrix},
\qquad \text{with } 0 < \alpha_2 < \pi.
\end{equation}
Since there is a infinite number of axes non-collinear with the
axis of the first rotation, we have infinitely many ways to generate SO(3) through
combinations of Higgs-family transformations.

In fact, the maximal invariance group [orthogonal to the gauged
hypercharge U(1)] of the SO(3)-symmetric scalar potential is
O(3), as noted in \cite{Ferreira:2009wh, Ivanov:2007de}.  This can
be seen either by examining the symmetry transformation
properties of the scalar fields
or via the transformations of the scalar field bilinears.
\Eq{eq37n} implies that the scalar potential is invariant with
respect to Higgs family transformations, $\varphi_i(x)\to
U_{ij}\varphi_j(x)$, for \textit{any choice} of the unitary matrix~$U$.
Since all potentially complex parameters
of the SO(3)-invariant scalar potential are zero,
\eq{eq37n} also implies that the scalar potential is invariant with
respect to GCP transformations, $\varphi_i(x)\to
X_{ij}\varphi_j^*(x^\prime)$, for \textit{any choice} of the unitary matrix $X$.
Mathematically, complex conjugation is isomorphic to the discrete group
$\mathbb{Z}_2$, and hence the relevant invariance group is
SO(3)$\otimes\mathbb{Z}_2\iso$ O(3).
In $K$ space, it immediately follows from \eq{eq37n} that
eqs.~\eqref{eq18} and \eqref{GCPsymm} are satisfied for all
proper and improper rotation matrices~$R$ and $\overline{R}$,
respectively.  In particular, using the results of Section 5.1,
we immediately conclude that the $K$ space symmetries of the
SO(3)-invariant scalar potential
can also be generated by a product of two non-collinear CP3 transformations,
$-R(\boldsymbol{\hat n}_1,\theta)$ and $-R(\boldsymbol{\hat n}_2,\theta)$, or
by a product of the Higgs family transformation
$R(\boldsymbol{\hat n}_1,\theta)$
and a CP3 transformation $-R(\boldsymbol{\hat n}_2,\theta)$, where
$\boldsymbol{\hat n}_1\boldsymbol{\times}\boldsymbol{\hat n}_2\neq 0$.
Thus, the full O(3) group
consisting of all proper and improper rotation matrices is
generated.\footnote{The reader
is cautioned that after the inclusion of the Yukawa sector,
the corresponding symmetry properties of each of these pairs of transformations
might be different from the pure scalar theory.}\\

\subsection{Identifying the THDM symmetry classes}
\label{sec:sum}

We have examined the constraints on the
scalar potential parameters for each of the six symmetry classes listed
in Table~I.  We can reverse the process and determine the symmetry
class given the scalar potential parameters.  We summarize the results of
Section 5.1 in Table II, which agrees, of course, with the 
basis-independent characterization of the symmetry classes 
given in~\cite{Ivanov:2005hg,Ivanov:2007de,Ferreira:2010hy}.
In particular, given a set of
constraints on $\boldsymbol{\xi}$, $\boldsymbol{\eta}$ and the
eigenvalues and eigenvectors of $E$,
one can uniquely determine the maximal symmetry class of the
scalar potential.
Note that $\boldsymbol{\xi}\boldsymbol{\times}\boldsymbol{\eta}=0$
for all symmetry classes, with the possible exception of CP1.
In the case of CP1, the only necessary conditions on the scalar potential
parameters are:\footnote{Using the results of Table II, it is
easy to check that $\boldsymbol{\xi}\boldsymbol{\cdot}\boldsymbol{\hat e}=
\boldsymbol{\eta}\boldsymbol{\cdot}\boldsymbol{\hat e}=0$ 
(where $\boldsymbol{\hat e}$ is one of the eigenvectors of $E$) must hold
for \textit{all} symmetry classes.  For the higher symmetry classes,
additional constraints must be satisfied as exhibited in Table II.}
\beq \label{cp1constraints}
\boldsymbol{\xi}\boldsymbol{\cdot}\boldsymbol{\hat e}_i=
\boldsymbol{\eta}\boldsymbol{\cdot}\boldsymbol{\hat e}_i=0\,,\qquad
(\boldsymbol{\xi},\boldsymbol{\eta})
\neq (0,0)\,,
\eeq
where $\boldsymbol{\hat e}_i$ is one of the three eigenvectors of $E$
(independently of the degeneracy of the eigenvalues of $E$).
When the scalar potential parameters are generic,
$\boldsymbol{\xi}\boldsymbol{\times}\boldsymbol{\eta}\neq 0$.
In this case the constraints given in \eq{cp1constraints} are
equivalent to the statement that
$\boldsymbol{\xi}\boldsymbol{\times}\boldsymbol{\eta}$ is an
eigenvector of $E$~\cite{Ivanov:2005hg}.\footnote{In this case, 
one can define $\axis$ as the unit
vector that points in the direction of
$\boldsymbol{\xi}\boldsymbol{\times}\boldsymbol{\eta}$, and
\eq{perpto} is automatically satisfied.}
On the other hand, if $\boldsymbol{\xi}\boldsymbol{\times}\boldsymbol{\eta}=0$,
then the symmetry class of the scalar potential will be CP1 if and
only if the constraints of the other symmetry classes
listed in Table II are not satisfied.  
For example, consider the case where one
eigenvalue of~$E$ is non-degenerate
and corresponds to the eigenvector 
$\boldsymbol{\hat e}_i$ in \eq{cp1constraints}, and the
other eigenvalue of $E$ is doubly-degenerate.
Then \eq{cp1constraints} necessarily implies that 
the collinear vectors $\boldsymbol{\xi}$
and $\boldsymbol{\eta}$ are also collinear with 
some linear combination of the other two eigenvectors
$\boldsymbol{\hat e}_j$ ($j\neq i$) of $E$, and the scalar potential therefore
exhibits either the $\mathbb{Z}_2$ or U(1) symmetry.  
If the eigenvalues of $E$ are non-degenerate, or if 
$\boldsymbol{\hat e}_i$ in \eq{cp1constraints} corresponds to 
a doubly-degenerate eigenvalue of $E$,
then it is possible that neither $\boldsymbol{\xi}$ nor $\boldsymbol{\eta}$ 
is an eigenvector of $E$, in which case the symmetry class 
of the scalar potential is CP1.

\begin{table}[t!]
\centering
\begin{tabular}{lll}
\hline
symmetry class \qquad\quad & constraints on $\boldsymbol{\xi}$ and
$\boldsymbol{\eta}$  \qquad\qquad\qquad\qquad\qquad\quad & constraints on $E$ \\
\hline
$\mathbb{Z}_2$ & $\boldsymbol{\xi}\times \boldsymbol{\hat e}
=\boldsymbol{\eta}\times \boldsymbol{\hat e}=0$; $(\boldsymbol{\xi},
\boldsymbol{\eta})\neq (0,0)$ &
eigenvalues are non-degenerate \\
& \qquad or \qquad & \qquad \qquad or \qquad\qquad \\
& $\boldsymbol{\xi}\boldsymbol{\times} \boldsymbol{\hat e^\prime}
=\boldsymbol{\eta}\boldsymbol{\times} \boldsymbol{\hat e^\prime}=0$; $(\boldsymbol{\xi},
\boldsymbol{\eta})\neq (0,0)$ &
$E$ possesses doubly-degenerate eigenvalues \\ \hline
U(1) & $\boldsymbol{\xi}\boldsymbol{\times}\boldsymbol{\hat e}
=\boldsymbol{\eta}\boldsymbol{\times}\boldsymbol{\hat e}=0$;
$(\boldsymbol{\xi},\boldsymbol{\eta})\neq (0,0)$ &
$E$ possesses doubly-degenerate eigenvalues  \\
&  \qquad or \qquad &  \qquad\qquad or \qquad\qquad \\
&  $\boldsymbol{\xi}\boldsymbol{\times}\boldsymbol{\eta}=0$;
$(\boldsymbol{\xi},\boldsymbol{\eta})
\neq (0,0)$ &
$E$ possesses triply-degenerate eigenvalues  ($E=\mu\id$) \\ \hline
SO(3) & $(\boldsymbol{\xi},\boldsymbol{\eta})=(0,0)$ &
$E$ possesses triply-degenerate eigenvalues ($E=\mu\id$)  \\ \hline
CP1 & $\boldsymbol{\xi}\boldsymbol{\times}\boldsymbol{\eta}$ is an
eigenvector of $E$ & eigenvalues are unconstrained \\
&  \qquad or \qquad &  \qquad\qquad or \qquad\qquad \\
&  $\boldsymbol{\xi}\boldsymbol{\times}\boldsymbol{\eta}=0$;
$\boldsymbol{\xi}\boldsymbol{\cdot}\boldsymbol{\hat e}=
\boldsymbol{\eta}\boldsymbol{\cdot}\boldsymbol{\hat e}=0$;
$(\boldsymbol{\xi},\boldsymbol{\eta})
\neq (0,0)$; &
eigenvalues are non-degenerate  \\
&
neither $\boldsymbol{\xi}$ nor $\boldsymbol{\eta}$ is an eigenvector of $E$
& \\ 
&  \qquad or \qquad &  \qquad\qquad or \qquad\qquad \\
&  $\boldsymbol{\xi}\boldsymbol{\times}\boldsymbol{\eta}=0$;
$\boldsymbol{\xi}\boldsymbol{\cdot}\boldsymbol{\hat e^\prime}=
\boldsymbol{\eta}\boldsymbol{\cdot}\boldsymbol{\hat e^\prime}=0$;
$(\boldsymbol{\xi},\boldsymbol{\eta})
\neq (0,0)$; &
$E$ possesses doubly-degenerate eigenvalues  \\
&
neither $\boldsymbol{\xi}$ nor $\boldsymbol{\eta}$ is an eigenvector of $E$
& \\ \hline
CP2 & $(\boldsymbol{\xi},\boldsymbol{\eta})=(0,0)$ &
eigenvalues are non-degenerate \\ \hline
CP3 &  $(\boldsymbol{\xi},\boldsymbol{\eta})=(0,0)$ &
$E$ possesses doubly-degenerate eigenvalues  \\
\hline
\end{tabular}
\caption{\label{tab2} \small
The symmetry classes and the corresponding constraints on the
scalar potential parameters.  The unit vector $\boldsymbol{\hat e}$ is 
one of the three eigenvectors of $E$ corresponding to a
non-degenerate eigenvalue of $E$, and the unit vector
$\boldsymbol{\hat e^\prime}$ is 
an eigenvector of $E$ corresponding to a
doubly-degenerate eigenvalue of $E$.
Note that if
$(\boldsymbol{\xi},\boldsymbol{\eta})=(0,0)$, then the eigenvectors of $E$
play no role in constraining the scalar potential parameters.
For the U(1) symmetry class,
$\boldsymbol{\hat e}$ is uniquely identified as the eigenvector of
$E$ corresponding to the non-degenerate eigenvalue (if it exists).
For the CP1 symmetry class, $\boldsymbol{\xi}$ and
$\boldsymbol{\eta}$ are necessarily orthogonal to one of the
eigenvectors of $E$.  If $\boldsymbol{\xi}$ and
$\boldsymbol{\eta}$ are collinear [and $(\boldsymbol{\xi},\boldsymbol{\eta})\neq(0,0)$],
then we must impose one additional condition to ensure that
the constraints relevant for the $\mathbb{Z}_2$ or U(1) symmetry classes
are \textit{not} satisfied.
}
\end{table}

Suppose that the scalar potential parameters satisfy the constraints
of a particular symmetry class.  By adding additional constraints,
it is often possible to promote the scalar potential to a class
with a larger symmetry.  That is, there is a hierarchy of symmetries
that can schematically be represented by the following chain:
\beq \label{hierarchy}
{\rm CP1}\,<\,\mathbb{Z}_2\,<\,\left\{\begin{array}{l}
{\rm U}(1)\\  {\rm CP2} \\ \end{array}\right\} \,<\,{\rm CP3}
\,<\,\rm{SO}(3)\,.
\eeq
The meaning of the above equation is the following.  A scalar
potential that exhibits a $\mathbb{Z}_2$ symmetry also must exhibit a
CP1 symmetry.  If the scalar potential exhibits a U(1)
symmetry, then it must also exhibit the symmetries that precede it in
the chain, and so on.  Note that the U(1) symmetry does not
imply a CP2 symmetry and vice versa.  However, a CP3-symmetric
potential must exhibit both the U(1) symmetry and the CP2
symmetry.
\Eq{hierarchy} is a consequence of the results of Table~\ref{tab2}.
For example, if we take a CP2-symmetric scalar potential and impose the
additional condition that one of the eigenvalues of $E$ is
doubly-degenerate, then we promote the symmetry to CP3.  Likewise, if we take
a U(1)-symmetric scalar potential (assuming $E\neq\mu\id$)
and then impose the
additional constraint $\boldsymbol{\xi}=\boldsymbol{\eta}=0$, then we
promote the symmetry to CP3.  The rest of the hierarchy of symmetries
can be deduced in a similar manner.

Note that there are cases in which the symmetry
is \textit{not} enhanced if the degeneracy of one of the eigenvalues
of $E$ is increased.  Indeed, these cases \textit{do not} correspond
to new symemtry classes beyond the six listed in
Table~\ref{tab2}.  This conclusion is a consequence of
a classification theorem, originally proved in~\cite{Ivanov:2007de} based on
a geometrical analysis,\footnote{Note that in~\cite{Ivanov:2007de} the derivation
of the symmetry classes assumes
{\em strong} stability of the potential, that is,
the quartic terms alone guarantee stability.
Here we consider the general case
including {\em weak} and {\em marginal} stability where
the stability of the potential is guaranteed
only after inclusion of the quadratic terms.
In fact, the classification of Table~\ref{tab2}
is valid even for unstable potentials.}
which states that the only possible symmetry classes of the THDM
are those listed in Table~\ref{tab2}.
This result can be checked by employing the renormalization
group equations (RGEs) for the scalar potential parameters.  In particular,
the constraints on the scalar potential parameters for a given
symmetry class must be respected at any scale and hence invariant under
renormalization group evolution.

The RGEs for the scalar potential
parameters in the general THDM can be found
in~\cite{Ma:2009ax,Haber:1993an,Grimus:2004yh,Ferreira:2009jb}.
One can check that all the parameter constraints exhibited in Table II are
invariant under renormalization group running.  However, if there
are further accidental degeneracies among the eigenvalues of $E$,
these degeneracies will \textit{not} in general be renormalization group
invariant.  As a simple example, consider a U(1)-invariant scalar
potential in a basis where $m_{12}^2=\lambda_5=\lambda_6=\lambda_7=0$.
That is, we have from~\eq{eq17}
\begin{equation}
\begin{split}
\label{eqrun}
&\eta_{00} =
\tfrac{1}{8}(\lambda_1 + \lambda_2) + \tfrac{1}{4}\lambda_3\;,\qquad
\tvec{\xi}=
\begin{pmatrix}
0 \\
0 \\
\xi_3=\tfrac{1}{2} (m_{11}^2-m_{22}^2)
\end{pmatrix}, \qquad
\tvec{\eta}=
\begin{pmatrix}
0\\
0\\
\eta_3=\tfrac{1}{8}(\lambda_1 - \lambda_2)
\end{pmatrix},\qquad\\
&E = \tfrac{1}{4} \diag(\lambda_4, \lambda_4, \tfrac{1}{2}(\lambda_1 + \lambda_2)-\lambda_3 )\;.
\end{split}
\end{equation}
In this case $\boldsymbol{\hat e}=(0,0,1)$, and the constraints listed
in Table II are clearly satisfied.  Now, suppose we add the following
additional constraint: $\lambda_1+\lambda_2=2(\lambda_3+\lambda_4)$.
In this case, the matrix $E$ possesses one triply-degenerate eigenvalue, and
$E=\mu\id$ with $\mu\equiv\lambda_4/4$.
However, this is not a new
symmetry class because the constraint that
$\lambda_1+\lambda_2=2(\lambda_3+\lambda_4)$ is not in general
renormalization group invariant.  Using the RGEs for the scalar
potential parameters with $\lambda_5=\lambda_6=\lambda_7=0$, we find:
\beq \label{rge1}
8\pi^2\beta_{\lambda_1+\lambda_2-2(\lambda_3+\lambda_4)}=
\bigl[\lambda_1+\lambda_2-2(\lambda_3+\lambda_4)\bigr]
\bigl[3(\lambda_1+\lambda_2)+2\lambda_4\bigr]+3(\lambda_1-\lambda_2)^2\,.
\eeq
Thus, if $\lambda_1\neq\lambda_2$, then the condition
 $\lambda_1+\lambda_2-2(\lambda_3+\lambda_4)=0$ is not stable under
 renormalization group running.  If in addition $\lambda_1=\lambda_2$ and
$m_{11}^2=m_{22}^2$, then the scalar potential is SO(3)-invariant,
corresponding to one of the other symmetry classes of Table II.
Note that it is sufficient to set $\lambda_1=\lambda_2$ in \eq{rge1} to
obtain a fixed point in the RGE at
$\lambda_1+\lambda_2=2(\lambda_3+\lambda_4)=0$.  This would correspond
to a softly broken SO(3)-invariant scalar potential, where the
soft-breaking is due to $m_{11}^2\neq m_{22}^2$.  Of course, this soft
symmetry breaking is invisible to the RGE running of the coefficients
of dimension-four terms of the Higgs Lagrangian.

In general, additional degeneracies in the eigenvalues of $E$ in Table
II are either unstable with respect to renormalization group running
or else correspond to a known symmetry class of enhanced symmetry that
is either exact or softly-broken by dimension-two terms of the scalar
potential.  Thus, we
conclude that no additional symmetry classes beyond those listed in
Table II exist, in agreement with the classification theorem
of \cite{Ivanov:2007de}.

\subsection{Symmetries in the $\boldsymbol{E}$-diagonal basis}
\label{sec:diag}

The parameters of the scalar potential given by \eq{eq5} are described in
the K space formalism by \eqs{eq-fourpar}{eq17}.  In particular, under U(2) basis
transformations of eq.~\eqref{eq11} $\boldsymbol{\xi}$ and $\boldsymbol{\eta}$ transform
as real three-vectors and $E$ transforms as a real Cartesian
second-rank symmetric tensor under SO(3), which is related to the U(2)
basis transformation of the scalar fields via \eq{Rabu}.
Starting in a generic basis, one can always transform to a special
basis in which $E$ is diagonal,
\beq \label{RD}
R_D ER^\trans_D\equiv E_D={\rm diag}(\mu_1,\mu_2,\mu_3)\,,
\eeq
where the $\mu_i$ are the eigenvalues of $E$ and $R_D\in$~SO(3).
We begin by assuming that the scalar potential parameters are generic.
In this case, the eigenvalues of $E$ are non-degenerate, and
$R_D$ is unique up to the ordering of its rows and/or columns, which
corresponds to a reordering of the diagonal elements of $E_D$.
The corresponding normalized eigenvectors $E$ are unique
(up to an irrelevant multiplicative phase) and will be denoted
$\boldsymbol{\hat e}_1$, $\boldsymbol{\hat e}_2$ and
$\boldsymbol{\hat e}_3$.
Working in the basis where $E$
is diagonal (henceforth called the $E$-diagonal basis)
is equivalent to expressing the matrix $E$ with respect to the
orthonormal basis $\{\boldsymbol{\hat e}_1\,,\,\boldsymbol{\hat e}_2\,,\,
\boldsymbol{\hat e}_3\}$.

Having chosen the $E$-diagonal basis, we now investigate the form of the
various THDM symmetry classes.   In each case, the symmetry transformation
is of the form:
\beqa
&&\tvec{K}(x) \to R \tvec{K}(x)\,,\qquad \text{for Higgs family symmetries}\,,\\
&&\tvec{K}(x) \to \overline{R} \tvec{K}(x')\,,\qquad \text{for GCP symmetries}\,,
\eeqa
where $R$ is a proper rotation and $\overline{R}$ is an improper rotation.
In Table III, we summarize the possible forms for $R$ and $\overline{R}$ and
identify the relevant symmetry class.  We also indicate the corresponding
constraints on the scalar potential parameters.  In some cases, the
imposition of the symmetry will require that the eigenvalues of $E$ exhibit
some degeneracy.  If the symmetry imposes no such constraint, we say
that the eigenvalues of $E$ are unconstrained (i.e.~non-degenerate for
generic choices of the parameters of the scalar potential).

\begin{table}[t!]
\begin{tabular}{lcl}
\hline
rotation matrices in the $E$-diagonal basis &  \quad symmetry class
\qquad & \quad constraints on scalar potential parameters \\
\hline
$\phm R(\boldsymbol{\hat e},\pi)$ & $\mathbb{Z}_2$ & $\boldsymbol{\xi}
\boldsymbol{\times} \boldsymbol{\hat e}=\boldsymbol{\eta}\boldsymbol{\times}\boldsymbol{\hat e}=0$;
eigenvalues of $E$ are unconstrained \\
$\phm R(\boldsymbol{\hat n},\pi)$, $\axis\neq\boldsymbol{\hat e}$
 & $\mathbb{Z}_2$ & $\boldsymbol{\xi}\boldsymbol{\times}\axis
 =\boldsymbol{\eta}\boldsymbol{\times} \axis=0$; eigenvalues of $E$ are degenerate \\
$\phm R(\boldsymbol{\hat e},\theta)$, $0<\theta<\pi$ & U(1) & $\boldsymbol{\xi}
\boldsymbol{\times} \boldsymbol{\hat e}=\boldsymbol{\eta}\boldsymbol{\times}\boldsymbol{\hat e}=0$;
eigenvalues of $E$  are doubly-degenerate \\
$\phm R(\boldsymbol{\hat n},\theta)$, $\axis\neq\boldsymbol{\hat e}$, $0<\theta<\pi$,  & U(1) & $\boldsymbol{\xi}\boldsymbol{\times}\axis=\boldsymbol{\eta}\times\axis=0$;
$E=\mu\id$  \\
$-R(\boldsymbol{\hat e},\pi)$ & CP1 &
$\boldsymbol{\xi}\boldsymbol{\cdot}\boldsymbol{\hat e}=\boldsymbol{\eta}\boldsymbol{\cdot}\boldsymbol{\hat e}=0$;
eigenvalues of $E$ are unconstrained \\
$-R(\boldsymbol{\hat n},\pi)$, $\axis\neq\boldsymbol{\hat e}$
 & CP1 &
$\boldsymbol{\xi}\boldsymbol{\cdot}\axis=\boldsymbol{\eta}\boldsymbol{\cdot}\axis=0$;
eigenvalues of $E$ are degenerate \\
$-R(\boldsymbol{\hat n},0)\equiv -\id$,
 & CP2 & $\boldsymbol{\xi}=\boldsymbol{\eta}=0$; eigenvalues of $E$ are unconstrained \\
$-R(\boldsymbol{\hat e},\theta)$, $0<\theta<\pi$ & CP3 & $\boldsymbol{\xi}=\boldsymbol{\eta}=0$; eigenvalues of $E$
are doubly-degenerate \\
$-R(\boldsymbol{\hat n},\theta)$, $\axis\neq\boldsymbol{\hat e}$ , $0<\theta<\pi$  &
CP3 [SO(3)] & $\boldsymbol{\xi}=\boldsymbol{\eta}=0$;
$E=\mu\id$  \\
\hline
\end{tabular}
\caption{\label{tab3} \small
The proper [and improper] rotation matrices $R$
[and $\overline{R}$] that generate the symmetry classes in $K$ space, where
the scalar fields are defined in the $E$-diagonal basis.
The unit vector $\boldsymbol{\hat e}$ is any one of the three eigenvectors
$\{\boldsymbol{\hat e}_1$, $\boldsymbol{\hat e}_2$, $\boldsymbol{\hat e}_3 \}$ of $E$.
In the cases of the $\mathbb{Z}_2$ and CP1 symmetry classes where $\axis\neq\boldsymbol{\hat e}_i$
(for $i=1, 2, 3$),
$E$ possesses doubly-degenerate eigenvalues if $\axis$ can be expressed as a linear combination
of two of the eigenvectors of $E$; otherwise $E=\mu\id$.
The last line of the table corresponds to a special case of the CP3 symmetry
class in which all three eigenvalues of $E$ are degenerate.  The resulting scalar potential
parameter constraints yield an SO(3)-invariant scalar potential.
}
\end{table}

We now discuss some of the salient points of Table III.  First, the constraints
on $\boldsymbol{\xi}$ and $\boldsymbol{\eta}$ are precisely the same as the ones given in
Table II.  These constraints are determined simply by identifying the
eigenvector of $R$ or $\overline{R}$ (if it exists) corresponding to the eigenvalue $+1$.
Indeed, when no eigenvalue $+1$ exists (as in the cases of CP2 and CP3), it follows that
$\boldsymbol{\xi}=\boldsymbol{\eta}=0$.  Second, in the cases of proper and improper
rotation matrices parameterized in terms of $\boldsymbol{\hat n}=\boldsymbol{\hat e}$,
where $\boldsymbol{\hat e}$ is one of the
eigenvectors of $E$, no constraints on the eigenvectors
of $E$ arise for the $\mathbb{Z}_2$, CP1 and CP2 symmetry classes (in which
case $\boldsymbol{\hat e}$ can be any one of the three eigenvectors of $E$).  For the U(1) and
CP3 symmetry classes, the eigenvalues of $R(\axis,\theta)$ and $\overline{R}\equiv -R(\axis,\theta)$ are non-degenerate,
and the corresponding eigenvectors orthogonal to $\boldsymbol{\hat n}$ are complex conjugates of
each other.  Since $E$ commutes with $R$, the eigenvectors of $R$ are also eigenvectors of $E$.
Finally, as the eigenvalues of the real symmetric matrix $E$ must be real, it follows
that the eigenvalue of $E$ corresponding to the complex conjugate pair of eigenvectors
is doubly-degenerate.

However, if $\boldsymbol{\hat n}\neq \boldsymbol{\hat e}_i$ (for $i=1, 2, 3$), then additional constraints on
the eigenvalues of $E$ are imposed.  Repeating the arguments of Section 5.1, we see that
$\boldsymbol{\hat n}$ must also be an eigenvector of $E$.  This can only be consistent with
$\boldsymbol{\hat n}\neq \boldsymbol{\hat e}_i$ if some of the eigenvalues of $E$ are degenerate,
in which case linear combinations of the $\boldsymbol{\hat e}_i$ within the degenerate subspace
are also eigenvectors of $E$.
For example, in the case of the $\mathbb{Z}_2$ or the CP1 symmetry classes,
if $\boldsymbol{\hat n}$ is a linear combination of two of the
eigenvectors of $E$, then \eq{eijn} yields a two-fold degeneracy in the
eigenvalues of $E$.  Likewise, if $\boldsymbol{\hat n}$ is a linear combination of
all three of the originally chosen eigenvectors of $E$, then \eq{eijn} yields a three-fold degeneracy in the
eigenvalues of $E$, which means that $E$ is proportional to the identity matrix (i.e.
$E=\mu\id$).  Likewise, in the case of the U(1) or the CP3 symmetry classes, if
$\axis\neq\boldsymbol{\hat e}_i$ (for $i=1, 2, 3$), then the eigenvectors of $R$ or $\overline{R}$,
denoted by $\boldsymbol{\hat m}$ and $\boldsymbol{\hat m}^*$ in Section 5.1, are
linear combinations of all three of the eigenvectors of $E$ [cf.~\eq{m1}].  Consequently,
\eq{eiipq} requires that all three eigenvalues of $E$ must be degenerate, i.e.
$E=\mu\id$.  Here is a simple example to illustrate the last point.  Suppose we work in
a coordinate system in which the
eigenvectors of $E$ are $\boldsymbol{\hat e}_1=(1,0,0)$,
$\boldsymbol{\hat e}_2=(0,1,0)$, and $\boldsymbol{\hat e}_3=(0,0,1)$.  One possible
choice for the axis of the rotation matrix $R(\axis,\theta)$ is
$\axis=\tfrac{1}{\sqrt{2}}(\boldsymbol{\hat e}_1+\boldsymbol{\hat e}_2)=\tfrac{1}{\sqrt{2}}(1,1,0)$.
In this case, we use \eq{m1} to obtain
\beq
\boldsymbol{\hat m}=\frac{1}{2}\biggl(\frac{1}{\sqrt{2}}\left(1+i\right)\,,\,
-\frac{1}{\sqrt{2}}\left(1+i\right)\,,\,-1+i\biggr)\,,
\eeq
where $\boldsymbol{\hat m}$ is the eigenvector of $R(\axis,\theta)$ corresponding to the
non-degenerate eigenvalue
$e^{i\theta}$ of $R(\axis,\theta)$ [under the assumption that $0<\theta<\pi$].
When we impose the symmetry constraint, $ER(\axis,\theta)=RE(\axis,\theta)$,
it follows that $\boldsymbol{\hat m}$ must also be an eigenvector of $E$.  This is
only possible if $E=\mu\id$, as indicated in the last row of Table III.  Any other choice of
$\axis\neq\boldsymbol{\hat e}_i$ (for $i=1, 2, 3$) would lead to the same conclusion.

To summarize, we have examined possible symmetry transformations in the $E$-diagonal basis,
in which the vectors $\boldsymbol{\xi}$ and $\boldsymbol{\eta}$ and the matrix $E$
are given with respect to the orthonormal basis constructed from the
eigenvectors of $E$,
namely $\{\boldsymbol{\hat e}_1\,,\boldsymbol{\hat e}_2\,,\boldsymbol{\hat e}_3\}$.
Applying a symmetry transformation $\pm R(\axis,\theta)$, where
$\axis\neq\boldsymbol{\hat e}_i$ (for $i=1, 2, 3$), yields the symmetry classes
$\mathbb{Z}_2$, U(1), CP1, CP2 and CP3, but with extra degeneracies among the
eigenvalues of $E$.
Nevertheless, these extra
degeneracies do not correspond in general to new enhanced symmetry classes as
argued in Section 5.2.  In particular, symmetry
transformations defined with respect to a specific basis choice can yield
constraints on the scalar potential parameters that are not stable with respect
to renormalization group running.

Of course, one can always redefine the scalar fields
of the tree-level scalar potential to achieve a particular choice of basis.
However, once higher order corrections are taken into account,
the energy scale of the scalar potential parameters becomes relevant.
For example, the diagonalization required to achieve the $E$-diagonal basis depends
on the renormalization scale $M$.   In general, the eigenvalues and the directions
of the eigenvectors of $E$ will \textit{not} remain fixed under
a change in the renormalization scale.  Thus, parameter constraints defined in some
specific basis relative to some scale $M$ will not automatically be preserved at
some other scale $M^\prime \neq M$.  When discussing restrictions on parameters
of the tree-level scalar potential due to possible symmetries, one has to check explicitly
that they are preserved by renormalization group evolution.

As an example, we consider the GCP transformation
\begin{equation}
\begin{pmatrix} \varphi_1(x) \\ \varphi_2(x) \end{pmatrix} \rightarrow
\begin{pmatrix} e^{-i\omega} & 0 \\ 0 & e^{i\omega} \end{pmatrix}\,
\begin{pmatrix} \varphi_1^*(x') \\ \varphi_2^*(x') \end{pmatrix}\,,
\label{transfj}
\end{equation}
for some fixed $\omega$ with $0 < \omega < \pi/2$, which corresponds
to an improper rotation in $K$ space given by
\begin{equation}
\overline{R}_\omega\,=\,
\begin{pmatrix}
\cos 2\omega & \phm \sin 2\omega & \phm 0 \\
\sin 2\omega & -\cos 2\omega & \phm 0 \\
 0 & 0 & 1
\end{pmatrix}.
\end{equation}
We require this to be a symmetry.
Using eqs.~(\ref{trcos}), (\ref{spcase}) and (\ref{indiv}), it follows that
\beq
\overline{R}_\omega\equiv -R(\axis,\pi)\,,\qquad \quad \text{with}~~\axis=(\sin\omega\,,\,-\cos\omega\,,\,0)\,,
\eeq
which means that $\overline{R}_\omega$ generates a CP1 transformation.  If we apply this transformation
in the $E$-diagonal basis, where $\boldsymbol{\hat e}_1=(1,0,0)$,
$\boldsymbol{\hat e}_2=(0,1,0)$, and $\boldsymbol{\hat e}_3=(0,0,1)$, then 
$\axis=\sin\omega\,\boldsymbol{\hat e}_1-\cos\omega\,\boldsymbol{\hat e}_2$ 
must also be an eigenvector of $E$.  Hence in the $E$-diagonal basis, 
$E={\rm diag}(\mu_1,\mu_1,\mu_3)$
possesses doubly-degenerate eigenvalues.

However, the existence of doubly-degenerate eigenvalues is not stable under renormalization
group evolution.  To verify this statement, we shall work in a real basis (i.e.~a basis in which
all scalar potential parameters exhibited in~\eq{eq5}
are real), which is stable under renormalization group running.
Using \eq{eq17}, the matrix $E$ then takes the form:
\beq \label{mat}
E= \frac{1}{4} \begin{pmatrix} \lambda_4+\lambda_5  & \quad 0  & \quad\lambda_6-\lambda_7 \\
0  & \quad \lambda_4-\lambda_5  & \quad 0 \\ \lambda_6-\lambda_7 & \quad 0 & \quad\tfrac{1}{2}(\lambda_1+\lambda_2)-\lambda_3
\end{pmatrix}\,,
\eeq
where $\lambda_5$, $\lambda_6$ and $\lambda_7$ are real.  In the $E$-diagonal basis,
$\lambda_6=\lambda_7$.  Applying the symmetry constraints, \eq{GCPsymm}, imposed by $\overline{R}_\omega$,
yields $\lambda_5=0$ and we see that $E$ possesses a doubly-degenerate eigenvalue, $\lambda_4$.
We first note that the diagonal form for $E$ is not preserved under renormalization group running.   In particular,
in the real basis,
\beq
8\pi^2\beta_{\lambda_6-\lambda_7}=(\lambda_6-\lambda_7)\left[3(\lambda_1+\lambda_2)+2\lambda_4+4\lambda_5\right]
+3(\lambda_6+\lambda_7)(\lambda_1-\lambda_2)\,.
\eeq
Since $\lambda_1\neq\lambda_2$ in general, we see that the diagonal form for $E$ is not stable.
However, to check whether the CP1 symmetry is enhanced, one must examine the eigenvalues of \eq{mat} to see whether
the doubly-degenerate eigenvalue is preserved or not under the renormalization group running.  A straightforward
computation shows that \eq{mat} has a doubly-degenerate eigenvalue if
\beq
D\equiv\lambda_5\left(\lambda_1+\lambda_2-2\lambda_3-2\lambda_4+2\lambda_5\right)-(\lambda_6-\lambda_7)^2=0\,.
\eeq
The beta function for $D$ is given by
\beqa
\beta_D&=&(\lambda_1+\lambda_2-2\lambda_3-2\lambda_4+2\lambda_5)\beta_{\lambda_5}+\lambda_5
\beta_{\lambda_1+\lambda_2-2\lambda_3-2\lambda_4+2\lambda_5}-2(\lambda_6-\lambda_7)\beta_{\lambda_6-\lambda_7}
\nonumber \\
&=& \lambda_5(\beta_{\lambda_1}+\beta_{\lambda_2}-2\beta_{\lambda_3}-2\beta_{\lambda_4})
+(\lambda_1+\lambda_2-2\lambda_3-2\lambda_4+4\lambda_5)\beta_{\lambda_5}-2(\lambda_6-\lambda_7)(\beta_{\lambda_6}
-\beta_{\lambda_7})\,.
\eeqa
Inserting the corresponding RGEs in the real
basis~\cite{Ma:2009ax,Haber:1993an,Grimus:2004yh,Ferreira:2009jb}
yields:
\beq \label{D}
8\pi^2\beta_D=4D(\lambda_1+\lambda_2+\lambda_3+2\lambda_4-\lambda_5)+3(\lambda_6+\lambda_7)^2
(\lambda_1+\lambda_2-2\lambda_3-2\lambda_4+2\lambda_5)
+3(\lambda_1-\lambda_2)\left[\lambda_5(\lambda_1-\lambda_2)
-2\lambda_6^2+2\lambda_7^2\right]\,.
\eeq
Indeed, as long as $\lambda_1=\lambda_2$ and $\lambda_7=-\lambda_6$ is not satisfied, we see that $D=0$
is not a fixed point of \eq{D}.  We recognize $\lambda_1=\lambda_2$ and $\lambda_7=-\lambda_6$ as the
constraints on the dimensionless parameters of the scalar potential in the
exceptional region of the parameter space (ERPS) identified in \cite{Ferreira:2009wh}.  When we impose
these ERPS conditions, the CP1 symmetry is promoted to a CP3 symmetry\footnote{In the $E$-diagonal basis,
the ERPS conditions on the dimensionless couplings of the scalar potential corresponds to
$\lambda_1=\lambda_2$ and $\lambda_5=\lambda_6=\lambda_7=0$.   One is free to make a change of
basis that simultaneously interchanges the rows and columns of $E$ while keeping $E$ diagonal.
In the new basis, $\lambda_1=\lambda_2$ and $\lambda_6=\lambda_7=0$ are maintained, while
$\lambda_5=0$ is transformed to $\lambda_5=\lambda_1-\lambda_3-\lambda_4$.  The latter reproduces
the conditions for a CP3-symmetric scalar potential given in Table I of~\cite{Ferreira:2009wh}.}
(modulo possible soft-symmetry breaking
squared-mass terms).  Outside of the ERPS, the double-degeneracy of the eigenvalue of $E$ is not
a renormalization group invariant, and the CP1 symmetry is not enhanced.

In Table III, when $\axis\neq\boldsymbol{\hat e}_i$ (for $i=1, 2, 3$), the enhanced degeneracies
in the eigenvalues of $E$ are not stable with respect to renormalization group running in
the cases of the $\mathbb{Z}_2$, U(1) and CP1 symmetry classes.  However, in the case of the
CP3 symmetry class, the constraints on the scalar potential parameters coincide with \eq{eq37n}, which
are the same constraints imposed by the SO(3)-invariant scalar potential.
In this case, the resulting parameter constraints are invariant with respect to
renormalization group running.  That is, \textit{the CP3 symmetry class with an enhanced
degeneracy of the eigenvalues of $E$ is promoted to the SO(3) symmetry class}.
In particular, it is possible
to generate an SO(3) Higgs family symmetry by applying any particular CP3
transformation with $\axis\neq\boldsymbol{\hat e}_i$ (for $i=1, 2, 3$)
in the $E$-diagonal basis.  This seems to be in conflict with the results
of Section 5.1, where it was shown that two non-collinear rotation
matrices are necessary in order
to generate the SO(3) symmetry class in a generic basis.  However, given that an
independent rotation matrix
$R_D$ is required [cf.~\eq{RD}] in order to define the $E$-diagonal basis,  it is perhaps not
surprising that the SO(3) symmetry class can be generated by applying a single (improper) rotation matrix
in the $E$-diagonal basis.

The classification of the possible THDM symmetries is best done in a
generic basis, where the structure of the various symmetry classes is
transparent, and the resulting constraints on the scalar potential
parameters can be obtained that are covariant with respect to basis
transformations.  Although it is possible to perform an analysis of
symmetry classes in a specific basis (the $E$-diagonal basis is a
convenient choice to consider for this purpose), the resulting
classification is complicated by enhanced symmetry constraints that
are typically not renormalization group invariant.  Such enhanced
symmetry points in the scalar potential parameter space are 
accidental in nature and are not indicative of any new symmetry
structures.

\section{Conclusions}

It is known that there are only six classes
of symmetry-constrained potentials in the THDM~\cite{Ivanov:2007de}.
Specific implementations were later explored in~\cite{Ferreira:2009wh},
using Higgs-family and generalized CP (GCP)
\textit{symmetries} of the THDM potential.
In this paper, we have presented an analysis of the symmetry
classes, which applies to completely general
scalar potentials. That is, our analysis applies to
scalar potentials that are stable
in the strong, weak, or marginal sense, or
even unstable.
Furthermore, we have
pursued the geometric $K$ space interpretation
of Higgs-family and GCP \textit{transformations}
\cite{Nagel:2004sw, Maniatis:2006fs};
the former are proper SO(3) rotations and the latter are improper
rotations of the field bilinears in $K$ space.
We have constructed the relevant rotations and shown
explicitly their effects on the Higgs scalar potential.
This combines some known results with new ones
into a unified scheme and sets the framework for our analysis.

The following results have thus been obtained.
We have clarified the relation of the classifications of GCP transformations
of~\cite{Maniatis:2007vn} and~\cite{Ferreira:2009wh}.
We have given a simple geometric proof relating SO(3) rotations to
two reflections through planes and improper rotations to one or three reflections through
planes in $K$ space. We have shown that any Higgs-family
transformation can be considered as a product of two CP1 transformations
and any GCP transformation is either a CP1 transformation or a product of
three CP1 transformations. Based on this result we have 
provided a geometric interpretation of the
surprising result presented in~\cite{Ferreira:2009wh} that all Higgs-family
and GCP symmetries in the THDM can be generated from suitable CP1 symmetries.

\acknowledgments{
The work of P.M.F. is supported in part by the Portuguese
\textit{Funda\c{c}\~{a}o para a Ci\^{e}ncia e a Tecnologia} (FCT)
under contract PTDC/FIS/70156/2006.  The work of H.E.H. is supported
in part by the U.S. Department of Energy, under grant
number DE-FG02-04ER41268 and in part by a Humboldt Research Award sponsored by
the Alexander von Humboldt Foundation.
The work of J.P.S. is funded by the Portuguese FCT through the projects
CERN/FP/109305/2009 and  U777-Plurianual,
and by the EU RTN project Marie Curie: MRTN-CT-2006-035505.
The work of M.M. is supported by
\textit{Deutsche Forschungsgemeinschaft project number NA296/5-1}.}
H.E.H. is grateful for the hospitality of the Bethe Center for
Theoretical Physics at the Physikalisches Institut der Universit\"at
Bonn, where part of this work was completed.

\clearpage
\appendix

\section{Properties of $\boldsymbol{3\times 3}$ proper rotation matrices}

In this Appendix, we review the properties of
$3\times 3$ proper rotation matrices.  Most of this
material is standard and can be found in \cite{tung,Rao,Altmann}.

The most general three-dimensional proper rotation
is represented by an SO(3) matrix, $R(\boldsymbol{\hat n},\theta)$,
whose form is uniquely specified by an axis of rotation,
$\axis$, and a rotation angle $\theta$.  Conventionally, a
positive rotation angle corresponds to a counterclockwise rotation.
The direction of the axis is determined by the right hand rule.
Simple geometrical considerations imply that:
\beqa
R(\boldsymbol{\hat n},\theta+2\pi k)&=&R(\boldsymbol{\hat n},\theta)\,,\qquad k=0,\pm 1\,\pm 2\,\ldots\,,\label{rnk} \\[8pt]
[R(\boldsymbol{\hat n},\theta)]^{-1}&=&R(\boldsymbol{\hat
  n},-\theta)=R(-\boldsymbol{\hat n},\theta)\,.\label{Rrel}
\eeqa
Combining these two results, it follows that
\beq \label{rnpt}
R(\boldsymbol{\hat n},2\pi-\theta)=R(-\boldsymbol{\hat n},\theta)\,,
\eeq
which
implies that any three-dimensional rotation can be described by a
counterclockwise rotation by $\theta$ about an arbitrary axis $\axis$,
where $0\leq\theta\leq\pi$.\footnote{In the convention
adopted here, the overall sign of $\axis$ is meaningful for
$\theta\neq 0$~mod~$\pi$.}
However, for $\theta=\pi$ \eq{rnpt} yields:
\beq \label{minusn}
R(\boldsymbol{\hat n},\pi)=R(-\boldsymbol{\hat n},\pi)\,,
\eeq
which means that
for the special case of $\theta=\pi$, $R(\boldsymbol{\hat n},\pi)$ and
$R(-\boldsymbol{\hat n},\pi)$ represent the \textit{same} rotation.
Finally, if $\theta=0$, then $R(\boldsymbol{\hat n},0)=\id$ is the identity
operator, independently of the direction of $\axis$.

An explicit form for a general three-dimensional proper rotation
is given by:
\beq \label{Rij}
R_{ij}(\boldsymbol{\hat n},\theta)=\delta_{ij}\cos\theta+n_i
n_j(1-\cos\theta)-\epsilon_{ijk}n_k\sin\theta\,.
\eeq
This result simplifies for the case of $\theta=\pi$,
\beq \label{hturn}
R_{ij}(\boldsymbol{\hat n},\pi)=2n_i n_j-\delta_{ij}\,.
\eeq
It is noteworthy that $R(\boldsymbol{\hat n},\theta)$ is
a symmetric matrix if and only if $\theta=0$~(mod~$\pi$).

Given the SO(3) matrix $R(\boldsymbol{\hat n},\theta)$, one can
determine the corresponding angle of rotation $\theta$ and axis of
rotation $\axis$.  By taking the trace of \eq{Rij}, we immediately obtain
\beq \label{trr}
{\rm Tr}~R(\boldsymbol{\hat n},\theta)=
1+2\cos\theta=1+e^{i\theta}+e^{-i\theta}\,.
\eeq
It immediately follows that
\beq \label{trcos}
\cos\theta=\tfrac{1}{2}({\rm Tr}~R-1)\,,
\eeq
which determines $\theta$ uniquely in the convention that
$0\leq\theta\leq\pi$.  The axis of rotation is given by:
\beq \label{ax}
\boldsymbol{\hat n}=\frac{1}{\sqrt{(3-{\rm Tr}~R)(1+{\rm Tr}~R)}}
\left(R_{32}-R_{23}\,,\,R_{13}-R_{31}\,,\,R_{21}-R_{12}\right)\,,\qquad
\text{for}~~\theta\neq 0~\text{mod}~\pi\,.
\eeq
For $\theta=0~(\text{mod}~\pi)$, $R(\boldsymbol{\hat n},\theta)$ is
symmetric and cannot be determined from \eq{ax}.
In this case, \eq{trcos}
determines whether $\cos\theta=+1$ or $\cos\theta=-1$.
If $\cos\theta=+1$, then $R_{ij}=\delta_{ij}$ and the axis $\axis$ is undefined.
If $\cos\theta=-1$, then \eq{hturn} determines
the direction of $\axis$ up to an overall sign.   That is,
\beqa
&&\axis~\text{is undetermined if}~\theta=0\,,\nonumber \\[8pt]
&&\axis=\left(\epsilon_1\sqrt{\half(1+R_{11})}\,,\,
\epsilon_2\sqrt{\half(1+R_{22})}\,,\,
\epsilon_3\sqrt{\half(1+R_{33})}\right)\,, \qquad {\rm if}~\theta=\pi\,,
\label{spcase}
\eeqa
where the individual signs $\epsilon_i=\pm 1$ are determined up to an
overall sign via\footnote{If $R_{ii}=-1$, where $i$ is a fixed index, then
$n_i=0$, in which case the corresponding $\epsilon_i$ is not well-defined.}
\beq \label{indiv}
\epsilon_i\epsilon_j=\frac{R_{ij}}{\sqrt{(1+R_{ii})(1+R_{jj})}}\,,\qquad
\text{for fixed}~i\neq j\,,\, R_{ii}\neq -1\,,\,R_{jj}\neq -1\,.
\eeq
The ambiguity of the overall sign of $\axis$ sign is not significant, since
$R(\axis,\pi)$ and $R(-\axis,\pi)$ represent the same
rotation as noted above [cf.~\eq{minusn}].

Since the axis of rotation $\axis$ is invariant under all rotations
$R(\boldsymbol{\hat n},\theta)$,
it follows that $\axis$ is an eigenvector of
$R(\boldsymbol{\hat n},\theta)$ with corresponding eigenvalue $+1$,
\beq \label{Rnn}
R(\boldsymbol{\hat n},\theta)\axis=\axis\,.
\eeq
Moreover, since $R(\boldsymbol{\hat n},\theta)$ is also a unitary matrix,
we know that its eigenvalues are pure phases.  Combining \eq{trr}
with ${\rm det}~R(\boldsymbol{\hat n},\theta)=1$,
it follows that the individual eigenvalues of
$R(\boldsymbol{\hat n},\theta)$ must be 1, $e^{i\theta}$ and
$e^{-i\theta}$.  For $\theta\neq 0$ (mod~$\pi$), these three
eigenvalues are non-degenerate.  In this case, it is convenient to
introduce normalized eigenvectors
$\boldsymbol{\hat m}$ and $\boldsymbol{\hat m}^*$,
corresponding to the eigenvalues $e^{i\theta}$ and $e^{-i\theta}$, respectively.
Note that whereas $\boldsymbol{\hat n}$ is a real vector corresponding to the
axis of rotation, $\boldsymbol{\hat m}$ and its complex conjugate
are complex vectors.  The vectors $\axis$, $\boldsymbol{\hat m}$
and $\boldsymbol{\hat m}^*$ are mutually
orthonormal with respect to the inner product of a complex vector space
(e.g., $\boldsymbol{\hat n}\cdot\boldsymbol{\hat m}=
\boldsymbol{\hat m}\cdot\boldsymbol{\hat m}=0$ and $\boldsymbol{\hat m}\cdot\boldsymbol{\hat m}^*=1$).
Explicitly, we have
\beq \label{m1}
\boldsymbol{\hat m}=\frac{1}{\sqrt{2}}\left(n_3+\frac{in_2(n_1-in_2)}{1+n_3}\,,\,
-in_3-\frac{in_1(n_1-in_2)}{1+n_3}\,,\,-(n_1-in_2)\right)\,,\qquad
\text{for}~\boldsymbol{\hat n}\neq -\boldsymbol{\hat z}\,,
\eeq
up to an overall phase factor that can be fixed by convention.
This form is not very useful as $n_3\to -1$.  However, we can use
\eq{Rrel} to obtain (up to an overall phase)
\beq \label{m2}
\boldsymbol{\hat m}=\frac{1}{\sqrt{2}}\left(-n_3-\frac{in_2(n_1+in_2)}{1-n_3}\,,\,
-in_3+\frac{in_1(n_1+in_2)}{1-n_3}\,,\,n_1+in_2\right)\,,\qquad
\text{for}~\boldsymbol{\hat n}\neq \boldsymbol{\hat z}\,.
\eeq
Clearly, the eigenvectors $\boldsymbol{\hat n}$,
$\boldsymbol{\hat m}$ and $\boldsymbol{\hat m}^*$
are \textit{independent} of the value of the rotation angle $\theta$.
In numerical work, it is convenient to use \eq{m1} for $n_3 \geq 0$ and \eq{m2} for
$n_3\leq 0$.   One can check that for $\boldsymbol{\hat n}\neq \pm\boldsymbol{\hat z}$, \eqs{m1}{m2}
are identical up to an irrelevant multiplicative overall phase.

In the case of $\theta=\pi$,
$R(\boldsymbol{\hat n},\pi)$ possesses one non-degenerate eigenvalue
$+1$ and two degenerate eigenvalues $-1$.  The former
is associated with the axis of rotation [cf.~\eq{Rnn}].  The eigenvectors
corresponding to the degenerate eigenvalues, denoted below by
$\boldsymbol{\hat m}_1$ and $\boldsymbol{\hat m}_2\equiv
\boldsymbol{\hat n\times\hat m}_1$ ,
can be chosen to be real and orthonormal.  A convenient
choice is:\footnote{In this case, any problem involving $n_3=-1$ can
be avoided simply by employing \eq{minusn}.}
\beq \label{hatm}
\boldsymbol{\hat m_1}=\left(n_3+\frac{n_2^2}{1+n_3}\,,\,
\frac{-n_1 n_2}{1+n_3} \,,\,-n_1\right)\,, \qquad\qquad
\boldsymbol{\hat m}_2=\left(\frac{-n_1 n_2}{1+n_3} \,,\,
n_3+\frac{n_1^2}{1+n_3}\,,\,-n_2\right)\,.
\eeq
However, any other orthonormal pair of vectors
constructed from linear combinations of
$\boldsymbol{\hat m}_1$ and $\boldsymbol{\hat m}_2$ would be
equally suitable.
Finally, in the case of $\theta=0$, $R(\boldsymbol{\hat n},\pi)=\id$
possesses three degenerate eigenvalues $+1$.

Finally, we prove an important result that is needed in the text.
Let $\widetilde{R}\in$~SO(3) such that
\beq \label{axisp}
\boldsymbol{\hat n}^\prime=\widetilde{R}\boldsymbol{\hat n}\,.
\eeq
Then
\beq \label{rtil}
R(\boldsymbol{\hat n}^\prime,\theta)
=\widetilde{R}R(\boldsymbol{\hat n},\theta)\widetilde{R}^\trans\,.
\eeq
To prove \eq{rtil}, we first note that
\beq
{\rm Tr}~R(\boldsymbol{\hat n}^\prime,\theta)=
 {\rm Tr}\bigl[\widetilde{R}R(\boldsymbol{\hat
   n},\theta)\widetilde{R}^\trans\bigr]
={\rm Tr}~R(\boldsymbol{\hat n},\theta)\,,
\eeq
using the cyclicity of the trace and $\widetilde{R}\widetilde{R}^\trans=\id$.
It follows from \eq{trr} that the angle of rotation of
$R(\boldsymbol{\hat n},\theta)$ and  $\widetilde{R}
R(\boldsymbol{\hat n},\theta)\widetilde{R}^\trans$ must be the same.
Next, we use \eq{Rnn} to determine the axis of rotation of
$\widetilde{R}R(\boldsymbol{\hat n},\theta) \widetilde{R}^\trans $,
\beq
\widetilde{R}R(\boldsymbol{\hat n},\theta)
\widetilde{R}^\trans(\widetilde{R}\boldsymbol{\hat n})
=\widetilde{R}R(\boldsymbol{\hat n},\theta)\boldsymbol{\hat n}
=\widetilde{R}\boldsymbol{\hat n}\,,
\eeq
which implies that $\widetilde{R}\boldsymbol{\hat n}$ is an
eigenvector of $\widetilde{R}R(\boldsymbol{\hat n},\theta)
\widetilde{R}^\trans$ with eigenvalue $+1$.  If $\theta\neq 0$, then
the eigenvalue $+1$ is non-degenerate, in which case
$\widetilde{R}\boldsymbol{\hat n}$
is the axis of rotation of $R(\boldsymbol{\hat n}^\prime,\theta)$, and
\eqs{axisp}{rtil} are confirmed.  If $\theta=0$, then \eq{rtil} is
trivially satisfied.

Specializing to the case of $\boldsymbol{\hat n}= \boldsymbol{\hat z}$,
it then follows that:
\beq \label{nz}
R(\boldsymbol{\hat n},\theta)=
\widetilde{R}R(\boldsymbol{\hat z},\theta)\widetilde{R}^\trans\,,
\qquad \text{where}~~\boldsymbol{\hat n}\equiv\widetilde{R}\boldsymbol{\hat z}
\,.
\eeq

%
%
%


\begin{thebibliography}{99}


\bibitem{pdg}
K.~Nakamura et al. [Particle Data Group],
  ``Review of particle physics,''
\mbox{J.\ Phys.\ G {\bf 37}, 075021 (2010)}.


\bibitem{Gunion:1989we}
  J.F.~Gunion, H.E.~Haber, G.L.~Kane and S.~Dawson,
  \textit{The Higgs Hunter's Guide}
  \mbox{(Westview Press, Boulder, CO, 2000)}.



\bibitem{Lee:1973iz}
  T.D.~Lee,
  ``A theory of spontaneous T violation,''
  \mbox{Phys.\ Rev.\  D {\bf 8}, 1226 (1973)}.

\bibitem{Lee:1974jb}
  T.D.~Lee,
  ``CP nonconservation and spontaneous symmetry breaking,''
  \mbox{Phys.\ Rept.\  {\bf 9}, 143 (1974)}.




\bibitem{Peccei:1977hh}
  R.D.~Peccei and H.R.~Quinn,
  ``CP conservation in the presence of instantons,''
  \mbox{Phys.\ Rev.\ Lett.\  {\bf 38}, 1440 (1977)}.

\bibitem{Peccei:1977ur}
  R.D.~Peccei and H.R.~Quinn,
  ``Constraints imposed by CP conservation in the presence of instantons,''
  \mbox{Phys.\ Rev.\  D {\bf 16}, 1791 (1977)}.



\bibitem{Fayet:1974fj}
P.~Fayet,
``A gauge theory of weak and electromagnetic interactions with spontaneous
parity breaking,''
\mbox{Nucl.\ Phys.\  B {\bf 78}, 14 (1974)};

\bibitem{Fayet:1974pd}
  P.~Fayet,
  ``Supergauge invariant extension of the Higgs mechanism and a model for the
  electron and its neutrino,''
  \mbox{Nucl.\ Phys.\  B {\bf 90}, 104 (1975)}.

\bibitem{Inoue:1982pi}
K.~Inoue, A.~Kakuto, H.~Komatsu and S.~Takeshita,
``Aspects of grand unified models with softly broken supersymmetry,''
\mbox{Prog.\ Theor.\ Phys.\  {\bf 68}, 927 (1982)
[Erratum-ibid.\  {\bf 70}, 330 (1983)]};
``Renormalization of supersymmetry breaking parameters revisited,''
\mbox{Prog.\ Theor.\ Phys.\  {\bf 71}, 413 (1984)}.

\bibitem{Flores:1982pr}
R.A.~Flores and M.~Sher,
``Higgs masses in the standard, multi-Higgs and supersymmetric Models,''
\mbox{Annals Phys.\  {\bf 148}, 95 (1983)}.

\bibitem{Gunion:1984yn}
  J.F.~Gunion and H.E.~Haber,
``Higgs bosons in supersymmetric models. I,''
\mbox{Nucl.\ Phys.\  B {\bf 272}, 1 (1986)}
\mbox{[Erratum-ibid.\  B {\bf 402}, 567 (1993)]}.




\bibitem{Deshpande:1977rw}
  N.G.~Deshpande and E.~Ma,
``Pattern Of Symmetry Breaking With Two Higgs Doublets,''
  Phys.\ Rev.\  D {\bf 18}, 2574 (1978).

\bibitem{Georgi:1978xz}
  H.~Georgi,
 ``A Model Of Soft CP Violation,''
  Hadronic J.\  {\bf 1}, 155 (1978).

\bibitem{Haber:1978jt}
  H.E.~Haber, G.L.~Kane and T.~Sterling,
``The Fermion Mass Scale And Possible Effects Of Higgs Bosons On Experimental
Observables,''
  Nucl.\ Phys.\  B {\bf 161}, 493 (1979).

\bibitem{Donoghue:1978cj}
  J.F.~Donoghue and L.~F.~Li,
``Properties Of Charged Higgs Bosons,''
  Phys.\ Rev.\  D {\bf 19}, 945 (1979).

\bibitem{Golowich:1978nh}
  E.~Golowich and T.C.~Yang,
``Charged Higgs Bosons And Decays Of Heavy Flavored Mesons,''
  Phys.\ Lett.\  B {\bf 80}, 245 (1979).

\bibitem{Hall:1981bc}
  L.J.~Hall and M.B.~Wise,
``Flavor Changing Higgs Boson Couplings,''
  Nucl.\ Phys.\  B {\bf 187}, 397 (1981).


\bibitem{Haber:1993an}
  H.E.~Haber and R.~Hempfling,
``The Renormalization group improved Higgs sector of the minimal
supersymmetric model,''
  Phys.\ Rev.\  D {\bf 48}, 4280 (1993)
  [arXiv:hep-ph/9307201].  We alert the reader to one typographical
  error in Appendix A of this paper.  In eq. (A2), replace
  $12\lambda_6^2$ with $12\lambda_7^2$.  Note that the renormalization
  equations given in Appendix A were derived for the CP-conserving
  THDM in which all scalar potential parameters are real.


\bibitem{Cvetic:1993cy}
  G.~Cvetic,
  ``CP violation in bosonic sector of SM with two Higgs doublets,''
  \mbox{Phys.\ Rev.\  D {\bf 48}}, 5280 (1993)
  \mbox{[arXiv:hep-ph/9309202]}.

\bibitem{Botella:1994cs}
  F.J.~Botella and J.P.~Silva,
  ``Jarlskog--like invariants for theories with scalars and fermions,''
  \mbox{Phys.\ Rev.\  D {\bf 51}, (1995) 3870}
  \mbox{[arXiv:hep-ph/9411288]}.

\bibitem{Lavoura:1994fv}
  L.~Lavoura and J.P.~Silva,
  ``Fundamental CP violating quantities in a SU(2) x U(1) model with many Higgs doublets,''
  \mbox{Phys.\ Rev.\  D {\bf 50}, 4619 (1994)}
  [arXiv:hep-ph/9404276].  



\bibitem{Lavoura:1994yu}
  L.~Lavoura,
 ``Signatures of discrete symmetries in the scalar sector,''
  Phys.\ Rev.\  D {\bf 50}, 7089 (1994)
  [arXiv:hep-ph/9405307].



\bibitem{Bernreuther:1998rx}
  W.~Bernreuther and O.~Nachtmann,
  ``Flavour dynamics with general scalar fields,''
  \mbox{Eur.\ Phys.\ J.\  C {\bf 9}, 319 (1999)}
  \mbox{[arXiv:hep-ph/9812259]}.


\bibitem{Ginzburg:2004vp}
  I.F.~Ginzburg and M.~Krawczyk,
  ``Symmetries of two Higgs doublet model and CP violation,''
  \mbox{Phys.\ Rev.\ D {\bf 72}, 115013 (2005)}
  \mbox{[arXiv:hep-ph/0408011]}.


\bibitem{Davidson:2005cw}
  S.~Davidson and H.E.~Haber,
 ``Basis-independent methods for the two-Higgs-doublet model,''
 \mbox{ Phys.\ Rev.\  D {\bf 72}, 035004 (2005)
  [Erratum-ibid.\  D {\bf 72}, 099902 (2005)]}
  \mbox{[arXiv:hep-ph/0504050]}.


\bibitem{Gunion:2005ja}
J.F.~Gunion and H.~E.~Haber,
``Conditions for CP-violation in the general two-Higgs-doublet model,''
\mbox{Phys.\ Rev.\ D {\bf 72}, 095002 (2005)}
\mbox{[arXiv:hep-ph/0506227]}.

\bibitem{Haber:2006ue}
  H.E.~Haber and D.~O'Neil,
``Basis-independent methods for the two-Higgs-doublet model. II: The
significance of $\tan\beta$,''
  \mbox{Phys.\ Rev.\  D {\bf 74}, 015018 (2006)}
  \mbox{[arXiv:hep-ph/0602242]}.


\bibitem{Barbieri:2005kf}
  R.~Barbieri and L.J.~Hall,
  ``Improved naturalness and the two Higgs doublet model,''
  \mbox{arXiv:hep-ph/0510243}.



\bibitem{Branco:2005em}
  G.C.~Branco, M.N.~Rebelo and J.I.~Silva-Marcos,
  ``CP-odd invariants in models with several Higgs doublets,''
  \mbox{Phys.\ Lett.\  B {\bf 614}, 187 (2005)}
  \mbox{[arXiv:hep-ph/0502118]}.

\bibitem{Barroso:2005sm}
  A.~Barroso, P.M.~Ferreira and R.~Santos,
  ``Charge and CP symmetry breaking in two Higgs doublet models,''
  \mbox{Phys.\ Lett.\  B {\bf 632}, 684 (2006)}
  \mbox{[arXiv:hep-ph/0507224]}.


\bibitem{Barbieri:2006dq}
  R.~Barbieri, L.J.~Hall and V.S.~Rychkov,
  ``Improved naturalness with a heavy Higgs: An alternative road to LHC physics,''
  \mbox{Phys.\ Rev.\  D {\bf 74}, 015007 (2006)}
  \mbox{[arXiv:hep-ph/0603188]}.


\bibitem{Fromme:2006cm}
  L.~Fromme, S.J.~Huber and M.~Seniuch,
  ``Baryogenesis in the two-Higgs doublet model,''
  \mbox{JHEP {\bf 0611}, 038 (2006)}
  \mbox{[arXiv:hep-ph/0605242]}.

\bibitem{Barroso:2007rr}
  A.~Barroso, P.M.~Ferreira and R.~Santos,
  ``Neutral minima in two-Higgs doublet models,''
  \mbox{Phys.\ Lett.\  B {\bf 652}, 181 (2007)}
  [arXiv:hep-ph/0702098].


\bibitem{Gerard:2007kn}
  J.M.~Gerard and M.~Herquet,
  ``A twisted custodial symmetry in the two-Higgs-doublet model,''
  \mbox{Phys.\ Rev.\ Lett.\  {\bf 98}, 251802 (2007)}
  \mbox{[arXiv:hep-ph/0703051]}.


\bibitem{Mahmoudi:2009zx}
  F.~Mahmoudi and O.~Stal,
  ``Flavor constraints on the two-Higgs-doublet model with general Yukawa couplings,''
  \mbox{[arXiv:0907.1791 [hep-ph]]}.

\bibitem{Ferreira:2010bm}
  P.M.~Ferreira and J.P.~Silva,
  ``A Two-Higgs Doublet Model With Remarkable CP Properties'',
accepted for publication in \mbox{Eur. Phys. J. C,}
  \mbox{[arXiv:1001.0574v2 [hep-ph]]}.



\bibitem{Velhinho:1994vh}
  J.~Velhinho, R.~Santos and A.~Barroso,
  ``Tree level vacuum stability in two-Higss doublet models,''
  \mbox{Phys.\ Lett.\  B {\bf 322}, 213 (1994)}.

\bibitem{Nagel:2004sw}
  F.~Nagel,
  ``New aspects of gauge-boson couplings and the Higgs sector,''
  Ph.~D. thesis (University of Heidelberg, 2004), available from
 \texttt{http://archiv.ub.uni-heidelberg.de/archiv/4803}.

\bibitem{Ivanov:2005hg}
  I.P.~Ivanov,
  ``Two-Higgs-doublet model from the group-theoretic perspective,''
  \mbox{Phys.\ Lett.\  {\bf B632}, 360 (2006)}
  \mbox{[hep-ph/0507132]}.

\bibitem{Maniatis:2006fs}
  M.~Maniatis, A.~von Manteuffel, O.~Nachtmann and F.~Nagel,
  ``Stability and symmetry breaking in the general two-Higgs-doublet model,''
  \mbox{Eur.\ Phys.\ J.\  C {\bf 48}, 805 (2006)}
  \mbox{[arXiv:hep-ph/0605184]}.

\bibitem{Maniatis:2006jd}
  M.~Maniatis, A.~von Manteuffel and O.~Nachtmann,
  ``Determining the global minimum of Higgs potentials via Groebner
   bases--applied to the NMSSM,''
  \mbox{Eur.\ Phys.\ J.\  C {\bf 49}, 1067 (2007)}
  \mbox{[arXiv:hep-ph/0608314]}.

\bibitem{Nishi:2006tg}
  C.C.~Nishi,
  ``CP violation conditions in $N$-Higgs-doublet potentials,''
  \mbox{Phys.\ Rev.\  D {\bf 74}, 036003 (2006)}
  \mbox{[Erratum-ibid.\  D {\bf 76}, 119901 (2007)]}
  \mbox{[arXiv:hep-ph/0605153]}.

\bibitem{Ivanov:2006yq}
  I.P.~Ivanov,
  ``Minkowski space structure of the Higgs potential in 2HDM,''
  \mbox{Phys.\ Rev.\  D {\bf 75}, 035001 (2007)}
  \mbox{[Erratum-ibid.\  D {\bf 76}, 039902 (2007)]}
  \mbox{[arXiv:hep-ph/0609018]}.

\bibitem{Ivanov:2007de}
  I.P.~Ivanov,
  ``Minkowski space structure of the Higgs potential in 2HDM. II. Minima, symmetries, and topology,''
  \mbox{Phys.\ Rev.\  {\bf D77}, 015017 (2008)}
  \mbox{[arXiv:0710.3490 [hep-ph]].}

\bibitem{Maniatis:2007vn}
  M.~Maniatis, A.~von Manteuffel and O.~Nachtmann,
  ``CP Violation in the general two-Higgs-doublet Model: a geometric view,''
  \mbox{Eur.\ Phys.\ J.\  C {\bf 57}, 719 (2008)}
  \mbox{[arXiv:0707.3344 [hep-ph]]}.

\bibitem{Maniatis:2007de}
  M.~Maniatis, A.~von Manteuffel and O.~Nachtmann,
  ``A new type of CP symmetry, family replication and fermion mass hierarchies,''
  \mbox{Eur.\ Phys.\ J.\  C {\bf 57}, 739 (2008)}
  \mbox{[arXiv:0711.3760 [hep-ph]]}.

\bibitem{Maniatis:2009vp}
  M.~Maniatis and O.~Nachtmann,
  ``On the phenomenology of a two-Higgs-doublet model with maximal CP symmetry at the LHC,''
  \mbox{JHEP {\bf 0905}, 028 (2009)}
  \mbox{[arXiv:0901.4341 [hep-ph]]}.

\bibitem{Ma:2009ax}
  E.~Ma and M.~Maniatis,
  ``Symbiotic Symmetries of the Two-Higgs-Doublet Model,''
  \mbox{Phys.\ Lett.\  B {\bf 683}, 33 (2010)}
  \mbox{[arXiv:0909.2855 [hep-ph]]}.


\bibitem{Maniatis:2009by}
  M.~Maniatis and O.~Nachtmann,
  ``On the phenomenology of a two-Higgs-doublet model with maximal CP symmetry
  at the LHC, II: radiative effects,''
  \mbox{[arXiv:0912.2727 [hep-ph]]}.


\bibitem{Ferreira:2009wh}
  P.M.~Ferreira, H.E.~Haber and J.P.~Silva,
  ``Generalized CP symmetries and special regions of parameter space in the
  two-Higgs-doublet model,''
  \mbox{Phys.\ Rev.\  D {\bf 79}, 116004 (2009)}
  \mbox{[arXiv:0902.1537 [hep-ph]]}.




\bibitem{Ecker:1981wv}
  G.~Ecker, W.~Grimus and W.~Konetschny,
  ``Quark mass matrices in left-right symmetric gauge theories,''
   \mbox{Nucl.\ Phys.\  B {\bf 191}, 465 (1981)}.

\bibitem{Ecker:1983hz}
  G.~Ecker, W.~Grimus and H.~Neufeld,
``Spontaneous CP violation in left-right symmetric gauge theories,''
  \mbox{Nucl.\ Phys.\  B {\bf 247}, 70 (1984)}.


\bibitem{Ecker:1987qp}
  G.~Ecker, W.~Grimus and H.~Neufeld,
  ``A standard form for generalized CP transformations,''
  \mbox{J.\ Phys.\ A  {\bf 20}, L807 (1987)}.


\bibitem{Neufeld:1987wa}
  H.~Neufeld, W.~Grimus and G.~Ecker,
 ``Generalized CP invariance, neutral flavor conservation and the
 structure of the mixing matrix,''
 \mbox{Int.\ J.\ Mod.\ Phys.\  A {\bf 3}, 603 (1988)}.


\bibitem{Branco:1999fs}
  G.C.~Branco, L.~Lavoura and J.P.~Silva,
  \textit{CP Violation} (Oxford University Press, Oxford, UK, 1999).


\bibitem{Nachtmann:1990ta}
  O.~Nachtmann,
  \textit{Elementary Particle Physics: Concepts And Phenomena}
\mbox{(Springer-Verlag, Berlin, Germany, 1990)}.

\bibitem{bigi}
I.I.~Bigi and A.I.~Sanda, \textit{CP Violation}, 2nd edition
(Cambridge University Press, Cambridge, UK, 2009).


\bibitem{Bernabeu:1986fc}
  J.~Bernabeu, G.C.~Branco and M.~Gronau,
  ``CP restrictions on quark mass matrices,''
  \mbox{Phys.\ Lett.\  B {\bf 169}, 243 (1986)}.



\bibitem{Dreiner:2008tw}
H.K.~Dreiner, H.E.~Haber and S.P.~Martin,
``Two-component spinor techniques and Feynman rules for quantum field theory
and supersymmetry,''
Phys.\ Rept.\  {\bf 494}, 1 (2010)
[arXiv:0812.1594 [hep-ph]].

\bibitem{horn}
R.A.~Horn and V.V.~Sergeichuk,
``Canonical forms for unitary congruence and $^*$congruence,''
Linear and Multilinear Algebra {\bf 57},
777 (2009)
\mbox{[arXiv:0710.1530 [math.RT]]}.

\bibitem{Grimus:1988qr}
  W.~Grimus, G.~Ecker,
``Basis transformations in generation space and a criterion for
the existence of standard forms for unitarily congruent matrices,''
J.\ Phys.\ A {\bf 21}, 2825 (1988).

\bibitem{Glashow:1976nt}
  S.L.~Glashow and S.~Weinberg,
  ``Natural conservation laws for neutral currents,''
  \mbox{Phys.\ Rev.\  D {\bf 15},  1958 (1977)}.

\bibitem{Paschos:1976ay}
  E.A.~Paschos,
  ``Diagonal neutral currents,''
  \mbox{Phys.\ Rev.\  D {\bf 15}, 1966 (1977)}.

\bibitem{Ferreira:2008zy}
  P.M.~Ferreira and J.P.~Silva,
  ``Discrete and continuous symmetries in multi-Higgs-doublet models,''
  \mbox{Phys.\ Rev.\  D {\bf 78}, 116007 (2008)}
  \mbox{[arXiv:0809.2788 [hep-ph]]}.

\bibitem{Ferreira:2010hy}
  P.M.~Ferreira, M.~Maniatis, O.~Nachtmann and J.~P.~Silva,
  ``CP properties of symmetry-constrained two-Higgs-doublet models,''
  \mbox{JHEP {\bf 1008}, 125 (2010)}
  \mbox{[arXiv:1004.3207 [hep-ph]].}


\bibitem{Grimus:2004yh}
  W.~Grimus and L.~Lavoura,
  ``Renormalization of the neutrino mass operators in the  multi-Higgs-doublet
  standard model,''
  Eur.\ Phys.\ J.\  C {\bf 39}, (2005) 219
  [arXiv:hep-ph/0409231].


\bibitem{Ferreira:2009jb}
  P.M.~Ferreira and D.R.T.~Jones,
  ``Bounds on scalar masses in two Higgs doublet models,''
  JHEP {\bf 0908} (2009) 069
  [arXiv:0903.2856 [hep-ph]].

\bibitem{tung}
W.-K.~Tung, \textit{Group Theory in Physics}
(World Scientific, Singapore, 1985).

\bibitem{Rao}
K.N.~Srinivasa Rao, \textit{The Rotation and Lorentz
Groups and their Representations for Physicists}
(Wiley Eastern Limited, New Delhi, 1988).

\bibitem{Altmann}
S.L.~Altmann,
\textit{Rotations, Quaternions, and Double Groups}
(Dover Publications, Inc., New York, NY, 2005).

\end{thebibliography}
\end{document}